\documentclass[fleqn,usenatbib]{mnras}

\usepackage[T1]{fontenc}
\usepackage{ae,aecompl}

\usepackage[english]{babel}        
\usepackage{microtype}             
\usepackage{xurl}                  

\usepackage{amssymb}
\usepackage{amsmath}
\usepackage{bm}

\usepackage{graphicx}
\usepackage{dblfloatfix}
\usepackage{booktabs}
\usepackage{threeparttable}
\usepackage{subcaption}
\usepackage{multirow}
\usepackage[font=small,labelfont=bf,tableposition=top]{caption}

\usepackage{color}
\usepackage[normalem]{ulem}
\usepackage{soul}
\usepackage{xcolor} 
\usepackage{ulem}   
\usepackage{cancel}  
\usepackage{natbib} 

\usepackage{lipsum} 

\usepackage{soul}
\soulregister\cite7
\soulregister\citep7
\soulregister\citet7
\soulregister\ref7

\newcommand{\fermi}{\textit{Fermi}-{\rm LAT}}

\newcommand{\chandra}{\textit{Chandra}}

\newcommand{\gray}{$\gamma$-ray~}

\newcommand{\xray}{X-ray~}
\newcommand{\xrays}{X-rays~}

\newcommand{\msun}{\mbox{$\rm M_\odot$}}

\newcommand{\nh}{\mbox{$N_{\rm H}$}}
\newcommand{\add}[1]{{#1}}   
\newcommand{\ad}[1]{{#1}}

\newcommand{\addmath}[1]{{#1}}
\newcommand{\remove}[1]{}

\newcommand{\addt}[1]{{#1}}
\newcommand{\addm}[1]{{#1}}
\newcommand{\rem}[1]{\textcolor{black}{}}

\newcommand{\adt}[1]{{#1}}
\newcommand{\adm}[1]{{#1}}
\newcommand{\re}[1]{\textcolor{black}{}}

\newcommand{\at}[1]{{#1}}
\newcommand{\am}[1]{{#1}}
\newcommand{\remo}[1]{\textcolor{black}{}}

\def\deg{\hbox{$^\circ$}}

\title[Jet-ejecta interaction within RG jets]{Could the interaction of jet and SN ejecta be the cause of \xray knots observed in \add{a} radio galaxy?}
\author[He et al.]{
Jia-Chun He$^{1}$, 
Xiao-Na Sun$^{1}$\thanks{xiaonasun@gxu.edu.cn}, 
Hao-Qiang Zhang$^{1}$, 
Yun-Feng Liang$^{1}$,
Hai-Ming Zhang$^{1}$,
\newauthor
Da-Bin Lin$^{1}$, 
En-Wei Liang$^{1}$\\ 
$^{1}$Guangxi Key Laboratory for Relativistic Astrophysics, School of Physics Science and Technology, Guangxi University, Nanning 530004, China\\
}

\usepackage{array} %
\let\oldtabular\tabular 
\let\endoldtabular\endtabular
\renewenvironment{tabular}{\oldtabular}{\endoldtabular} %
\begin{document}
\label{firstpage}
\pagerange{\pageref{firstpage}--\pageref{lastpage}}
\maketitle
\begin{abstract}
\ad{We investigate the interaction between relativistic jets and supernova (SN) ejecta as a potential origin of X-ray knots in radio galaxies, employing knot A in M 87 as a test case. 
By modeling the dynamical evolution of the interaction, we evaluate this scenario based on particle acceleration efficiency and spatial morphology. 
Our modeling indicates that the ejecta shock expands to only $\sim 30$ pc, which is inconsistent with the observed spatial scale of knot A ($\sim 60$ pc). 
In contrast, the jet shock can successfully reproduce the observed scale after approximately $3000$ yr\adt{, with the ejecta being accelerated to a bulk velocity of $\beta_{\rm ej}\approx 0.43$.} 
We fit the multi-wavelength spectral energy distribution (SED) using a one-zone leptonic framework, attributing the X-rays to synchrotron radiation from electrons accelerated up to $\sim1$ PeV at the jet shock. 
The derived magnetic field is approximately $70\ \mu\rm G$ \at{in the SN ejecta rest frame}, which is significantly below the equipartition value\re{, indicating that the dissipated energy is channeled primarily into particle acceleration rather than magnetic field amplification}. 
Protons \addt{may} be accelerated up to $\sim$ EeV, supporting the hypothesis that the jets of radio galaxies (RGs) may be the potential site for ultra-high-energy cosmic-ray (UHECR) acceleration within the framework of the jet-ejecta interaction. 
} 

\end{abstract}

\begin{keywords}
galaxies: jet -- \xrays: galaxies -- acceleration of particles -- radiation mechanism: non-thermal
\end{keywords}



\section{Introduction}
Radio galaxies (RGs) are a subclass of active galactic nuclei (AGNs), characterized by relativistic jets launched from the radio core of their host galaxies that extend from kiloparsec (kpc) to megaparsec (Mpc) scales.
Bright knots are common dominant structures within RG jets across radio, optical, and X-ray wavelengths, as seen in M 87 \citep{Perlman2005ApJ} and 3C 273 \citep{Jester2006, Jester2007}. 
The radio and optical emissions from RG jets are generally attributed to synchrotron radiation of electrons. However, the origin of the extended X-ray emission is still unclear \citep{Harris2006}. 
It has been proposed that inverse Compton scattering of cosmic microwave background photons by low-energy electrons (tens of MeV) (IC/CMB) could account for the X-ray emission \citep{Zhang2009,Breiding2017, Zhang2018a, Zhang2018b}. Nevertheless, this scenario is challenged by recent polarimetry observations and \gray observations (see also \citet{Georganopoulos2016}, for a review). 
The detection of extended TeV emission from the large-scale jet of Centaurus A by the High Energy Stereoscopic System (H.E.S.S.) supports a synchrotron origin for the X-ray emission, implying the presence of high-energy electrons with energies up to $\sim 100$ TeV \citep{HESS20}.
Corresponding tests have been proposed to explain the acceleration of these ultra-high-energy electrons. These include the stochastic (classical second-order Fermi) acceleration in the relativistic jet of Pictor A \citep{Fan2008ApJ}, shear (another type of second-order Fermi) acceleration in the knots of 3C 273 \citep{Rieger2019, Rieger2022a, Wang2021MNRAS, Wang2023MNRAS, He2023MNRAS}, and shock (classical first-order Fermi) acceleration in M 87 jet \citep{Sahayanathan2008MNRAS}. 
Shock acceleration, widely regarded as the best-understood mechanism, is commonly adopted as the primary acceleration process in the RG jets, motivated by its successful application in numerous other astrophysical scenarios.

Shocks can naturally form at the interface between jets and dense obstacles, such as the SN ejecta produced by stellar evolution within the jet environment \citep{Blandford1979ApL, Vieyro2019A&A, Bosch-Ramon2023A&A}. 
Jet-star interactions are expected to be frequent, with the number of such events potentially reaching up to $10^{8}$ within the first kiloparsec of the jet \citep{Vieyro2017A&A, Torres2019A&A, Fichet2025A&A}.  
This high frequency of interactions implies the presence of a significant stellar population within the jet environment. 
Stars reaching the end of their lives within the jet may undergo SN explosions. 
The subsequent expansion of SN ejecta inside the jet forms a massive, long-lived obstacle.
The dynamical interaction between the jet and the SN ejecta (hereafter jet-ejecta interaction) is a complex, multi-stage process. 
The collision between the jet and the SN ejecta drives a shock system comprising an ejecta shock that heats the ejecta and a jet shock that decelerates the flow. 
In the early phase, the ejecta shock \at{can be}\remo{is} a highly efficient particle accelerator, particularly for seeding and pre-accelerating heavy nuclei from the ejecta itself into the relativistic regime \citep{Bosch-Ramon2023A&A}. 
While present from the onset, the jet shock is expected to become the dominant and persistent accelerator during later stages, once the SN ejecta has been fully shocked, inflated, and mixed with the jet flow \citep{Vieyro2019A&A}.
In the later phase, the jet shock acts as a persistent accelerator, capable of dissipating jet energy into the shocked and mixed ejecta material. 
\ad{The scenario of SN exploding within extragalactic jets, along with its theoretical analysis, has been explored in several studies }\citep{Blandford1979ApL, Fedorenko1996A&A, Bednarek1999ptep, Vieyro2019A&A}. 
For example, \citet{Blandford1979ApL} established the theoretical foundation and proposed that jet-ejecta interaction could cause the knots observed in the relativistic jet of M 87. 
\citet{Vieyro2019A&A} focused on dynamical evolution and multi-wavelength emission and developed a dynamical model of supernova remnants (SNRs) interacting with relativistic jets and computed their multi-wavelength non-thermal emission. 
\citet{Bosch-Ramon2023A&A} has shown that jet-ejecta interactions can energize both protons and heavy nuclei up to $\sim$ EeV.
\ad{The dynamical evolution of the jet-ejecta interaction has been quantitatively revealed by recent high-resolution relativistic hydrodynamical (RHD) simulations \citep{Longo2025A&A}.}
However, the \ad{jet-ejecta interaction} scenario has not yet been applied to model the multi-wavelength spectral energy distributions (SEDs) of a specific knot observed in RG jets.

We aim to investigate whether the jet-ejecta interaction can produce the observed \xray emission in RG jets by examining two critical diagnostics, including particle acceleration efficiency and spatial extent of the \xray knots. 
In Section \ref{section:Model}, we describe the dynamic evolution of the ejecta shock and jet shock, particle acceleration, and the solution of the particle distribution function in the framework of the jet-ejecta interaction.
In Section \ref{section: application}, we apply this framework to analyze the relativistic jet in M 87. 
The summary and discussion are given in Section \ref{section:conclusion}.


\section{Model} \label{section:Model} 

\subsection{The dynamical evolution of shock system} \label{subsenction: evolution} 
\begin{figure}
    \centering
    \includegraphics[scale=0.25]{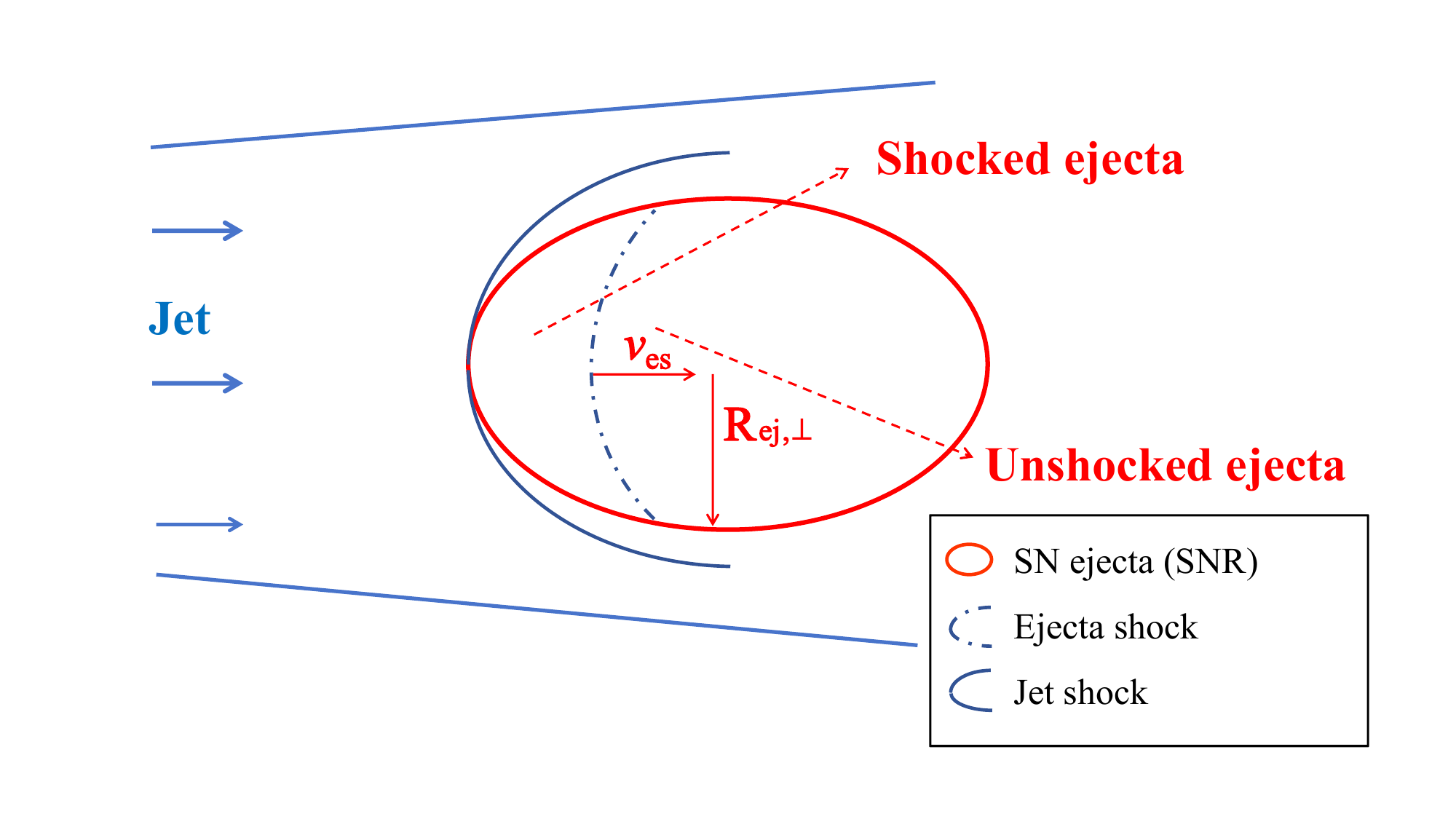}
    \caption{The schematic illustration of the jet-ejecta scenarios. The solid and dashed arcs denote the jet shock and the ejecta shock, respectively. The red circle is the profile of SN ejecta. } 
    \label{fig: model} 
\end{figure}

We model a SN explosion occurring within a relativistic jet, as illustrated in Figure \ref{fig: model}, following recent hydrodynamic studies \at{\citep{Vieyro2019A&A, Longo2025A&A}}\remo{\citep{Vieyro2019A&A, Bosch-Ramon2023A&A}}. 
This scenario involves an SN with a total ejecta mass of $M_{\rm ej}$ (typically 2 $\msun$, \addt{between the mass released in Type Ia and in core-collapse SNe} \citep{Dayal2018PhR}).
During the early stages of the explosion, the pressure exerted by the relativistic jet is lower than the SN ejecta pressure. 
Subsequently, the pressure of the SN ejecta drops due to its expansion. 

When the jet ram pressure balances that of the expanding ejecta, the radius of the SN ejecta is as follows \citep{Bosch-Ramon2023A&A}, 
\begin{equation}
R_{\rm ej, 0} \approx 13.4 \times{\left(\frac{E_{\rm SN}}{10^{51}\ \rm erg}\right)}^{1/3}{\left(\frac{L_{\rm j}}{10^{44}\ \rm erg\ s^{-1}}\right)}^{-1/3} \rm pc, \\
\label{eq:R_ej}
\end{equation}
which is valid for a jet radius of $\sim 100$ pc. Here $E_{\rm SN} = 10^{51}\ \rm erg$ and $L_{\rm j}$ denote the SN explosion energy and jet power, respectively. 
Meanwhile, an ejecta shock driven by the jet ram pressure propagates into the unshocked SN ejecta with a velocity of \adt{$v_{\rm es}(t) \approx v_{0}(1+\hat {x})^{3/2}$}\re{$v_{\rm s}(t) \approx v_{0}(1+\hat {x})^{3/2}$}, where $v_{0} = \sqrt{10E_{\rm SN}/3M_{\rm ej}}$ is the initial expansion velocity in the laboratory frame and $\hat {x} = v_{0}t/R_{\rm ej, 0}$ is the dimensionless time.
The ejecta shock will traverse the entire ejecta on a timescale $t_{\rm max}$ (maximum shock age), which can be derived as follows \citep{Bosch-Ramon2023A&A}
\begin{equation}
\adm{\int_0^{t_{\rm max}}v_{\rm es}(t)dt\approx 2R_{\rm ej,0}+v_{0}t_{\rm max}. }
\label{eq: shock length}
\end{equation}

The jet is also shocked when it collides with the SN ejecta. 
The timescale of the entire interaction process is $\sim 10^{4}$ yr. 
The energy dissipated by the jet shock during the ejecta-shocking phase is a factor of \adt{$v_{\rm j}/v_{\rm es}\sim 10$}\re{$v_{\rm j}/v_{\rm s}\sim 10$} higher than that at the ejecta shock (assuming a jet velocity of $c$) \citep{Bosch-Ramon2023A&A}, where $v_{\rm j}$ is the velocity of jet. 
The total energy dissipated over the entire interaction lifetime ($\sim10^{4}$ yr) is substantially larger than that dissipated during the initial ejecta-shocking phase, as the interaction lasts much longer than that phase, even if it weakens with time. 
The RHD simulations show that the expansion is enhanced as more energy from the jet is converted into kinetic and internal energy of the SN ejecta. 
The expanding radius of the unshocked ejecta is governed by its internal pressure and can be modelled as $R_{\rm ej}(t) \approx R_{\rm ej, 0} + \int_0^{t}c_{\rm s}dt^{\prime},$ where $c_{\rm s} \sim 0.01$ pc/yr.
The SN ejecta eventually covers most of the jet cross-section \citep{Longo2025A&A}. 
\addt{This leads} to the jet shock becoming a persistent, efficient, and large-scale site for energy dissipation, where the kinetic energy of the jet is converted into internal energy, thereby providing the sustained acceleration for non-thermal particles. 
The SN ejecta momentum increases due to the acceleration imparted by the jet along the direction of motion.
\addt{The bulk velocity of the SN ejecta can be estimated by} \citep{Barkov2012ApJ, Vieyro2019A&A}
\begin{equation}
    \addmath{\frac{d\beta_{\mathrm{ej}}}{dt} = \frac{L_{\rm j}}{M_{\mathrm{ej}} c^2} 
    \left( \frac{R_{\mathrm{ej}}}{R_{\rm j}} \right)^2 
    \left(1 - \frac{\beta_{\mathrm{ej}}}{\beta_{\rm j}}\right)^2
    (1-\beta_{\mathrm{ej}}^2)^{1/2}}, \label{eq: proper motion}
\end{equation} 
\ad{where $\beta_{\rm ej}$ is the velocity of SN ejecta in units of the speed of light, $\beta_{\rm j} = v_{\rm j}/c$. 
} 



\subsection{Particle acceleration \adt{and radiation}} \label{acceleration} 
The shock structures generated during the jet-ejecta interaction are potential sites for efficient particle acceleration.
We present the acceleration framework applicable to the shock system (i.e., either the ejecta shock or the jet shock). 
We focus on electron acceleration, as the observed radio-to-\xray emission from the knots is considered predominantly leptonic in origin (dominated by synchrotron and IC/CMB processes). 

We refer to the laboratory frame and the SN ejecta rest frame as $K$ and $K^{\prime}$, \adt{the primed and unprimed quantities denote those measured in $K^{\prime}$ and $K$,} respectively. 
\adt{The mean magnetic field strength in $K^{\prime}$ is given by}
\begin{align}
\adm{B^{\prime}_{\rm s}(d) = \frac{2}{\Gamma_{\rm ej}d}\sqrt{\frac{\zeta_{\rm ej}L_{\rm j}}{\theta^{2}_{\rm j} c}},} \label{eq: magnetic field} 
\end{align}
\adt{where $d = 1$ kpc is the distance to the black hole, $\zeta_{\rm ej}$ is the fraction of the total magnetic energy density to jet energy density, $\theta_{\rm j}$ is the jet inclination angle with respect to the line of sight, $\Gamma_{\rm ej} = (1-\beta_{\rm ej}^{2})^{-\frac{1}{2}}$ is the Lorentz factor of the SN ejecta.}

\adt{
The discrepancy between the radio-to-optical and X-ray spectral indices ($\alpha_{\rm RO}$ and $\alpha_{\rm X}$), as derived in Appendix \ref{appendix_data} and summarized in Table \ref{tab:Xray_data}, indicates that a single electron population cannot adequately reproduce the broadband SED. 
We adopt, for simplicity, a two-component model for the electron distribution. 
}
\re{Adopting a phenomenological approach, here we assume that both two electron populations are energized by the jet-ejecta interaction.}
In $K^{\prime}$, the radio-to-optical emission is attributed to the synchrotron radiation from a low-energy electron (LE) population. 
\adt{
We employ a phenomenological distribution to characterize this component, }
\re{which follows an exponential cut-off power-law spectrum described by $N^{\prime}_{\rm LE}(E^{\prime})=A_{1}(E^{\prime}/E^{\prime}_{0})^{-\alpha_{1}}{\rm exp}[-(E^{\prime}/E^{\prime}_{\rm cut1})^{2}]$ for $E^{\prime}>E^{\prime}_{\rm min1}$,} 
\begin{equation}
\adm{N^{\prime}_{\rm LE}(E^{\prime})=N^{\prime}_{\rm 0,LE}\left(\frac{E^{\prime}}{E^{\prime}_{\rm 0,LE}}\right)^{-\alpha_{\rm LE}}{\rm exp}\left[-\left(\frac{E^{\prime}}{E^{\prime}_{\rm cut, LE}}\right)^{2}\right],} \label{eq: LE}
\end{equation}
\adt{where $\alpha_{\rm LE}$ is the spectral index, \re{$A_{1}$} \adt{$N^{\prime}_{\rm 0,LE}$} denotes the normalization constant, $E^{\prime}_{\rm cut, LE}$ is the maximum energy of the LEs. 
We set the minimum energy of the LE population to 1 GeV.
}

\re{We consider the high-energy electron (HE) component to be responsible for the \xray emission. }
\adt{The \xray emission is attributed to a high-energy electron (HE) population. 
In contrast to the LE component, the extremely high energy of these electrons results in rapid radiative cooling. 
Their distribution is governed by the instantaneous balance between continuous injection and radiative cooling. 
The injection spectrum is defined as a power-law with an exponential cutoff,} 
\begin{align}
\adm{Q^{\prime}_{\rm HE}(E^{\prime}) = Q^{\prime}_{\rm 0,HE} {\left(\frac{E^{\prime}}{E^{\prime}_{\rm 0,HE}}\right)}^{-\alpha_{\rm HE}} {\rm exp}\left[-\left(\frac{E^{\prime}}{E^{\prime}_{\rm e, max}}\right)^{2}\right],} \label{eq: injected rate}
\end{align}
\adt{where $Q^{\prime}_{\rm 0,HE}$ and $\alpha_{\rm HE}$ are the normalization constant and the spectral index, respectively. $E^{\prime}_{\rm e,max}$ is the maximum energy of the HE population.
The resulting steady-state electron distribution $N^{\prime}_{\rm HE}$ in $K^{\prime}$ is obtained by solving the transport equation for a one-zone model,}
\begin{align}
\adm{N^{\prime}_{\rm HE}(E^{\prime}) = \frac{1}{\dot{E}^{\prime}} \int^{E^{\prime}_{\rm e, max}}_{E^{\prime}}Q^{\prime}_{\rm HE}(E^{*})dE^{*},} \label{eq: steady state}
\end{align}
\adt{where $\dot{E}^{\prime}$ represents the total energy loss rate, which is dominated by the synchrotron emission and the inverse Compton process. }

\re{During the acceleration process, electron cooling is dominated by synchrotron emission and IC/CMB process. }
At a shock front characterized by the magnetic field $B^{\prime}_{\rm s}$, the characteristic acceleration timescale \addt{in $K^{\prime}$} is given by \citep{Drury1983RPPh} 
\begin{equation}
\addm{t_{\rm acc}^{\prime}(E^{\prime}, B^{\prime}_{\rm s}) = \frac{E^{\prime}}{\eta ecB^{\prime}_{\rm s}}.} \label{eq: t_ac}
\end{equation}
The energetic electrons suffer radiation losses, with a characteristic cooling time \addt{in $K^{\prime}$ \citep{Rybicki1979rpa..book.....R},}
\begin{equation}
\adm{t_{\rm syn}^{\prime}(E^{\prime}, B^{\prime}_{\rm s}) = \frac{6(m_{\rm e}c^{2})^{2}\pi}{\sigma_{\rm T}cB_{\rm s}^{\prime 2}E^{\prime}},} \label{eq: t_sy}
\end{equation}
\begin{equation}
\adm{t_{\rm IC}^{\prime}(E^{\prime}) = \frac{3(m_{\rm e}c^{2})^{2}}{4c\sigma_{\rm T}E^{\prime} U^{\prime}_{\rm ph}},} \label{eq: t_IC}
\end{equation}
\begin{equation}
\adm{t_{\rm cool}^{\prime}(E^{\prime}, B^{\prime}_{\rm s}) = \frac{1}{t^{\prime}_{\rm syn}(E^{\prime}, B^{\prime}_{\rm s})^{-1} + t^{\prime}_{\rm IC}(E^{\prime})^{-1}},} \label{eq: t_cool}
\end{equation}
where $\sigma_{\rm T}$ is the Thomson scattering cross-section, \at{$U^{\prime}_{\rm ph} \approx 4.18 \times 10^{-13}(1+z)^{4}\Gamma_{\rm ej}^{2}\ \rm erg\ cm^{-3}$}\remo{$U^{\prime}_{\rm ph} = 4.18 \times 10^{-13}(1+z)^{4}\Gamma_{\rm ej}^{2}\ \rm erg\ cm^{-3}$} is the CMB energy density in $K^{\prime}$ \citep{Dermer2002ApJ}. 
\at{The Thomson cooling approximation is adopted, which is reasonable given the low energy of the target photons.}
The acceleration efficiency factor, $\eta$, is determined by the instantaneous shock velocity as $\eta = (v^{\prime}(t^{\prime})/c/2\pi)^2$, where $v^{\prime}(t^{\prime})$ is either $v^{\prime}_{\rm es}(t^{\prime})$ \at{(ejecta shock)} or $v^{\prime}_{\rm js}(t^{\prime})$ \at{(jet shock)}. 
In $K^{\prime}$, the velocities of the two shocks are derived from their respective lab-frame ($K$) values via Lorentz transformations. 
The time coordinate transforms as $t^{\prime}=t/\Gamma_{\rm ej}$. 
The velocity of the ejecta shock in $K^{\prime}$ is obtained by transforming its lab-frame velocity $v^{\prime}_{\rm es}(t^{\prime}) = (v_{\rm es}(t^{\prime})-\beta_{\rm ej}c)/(1-v_{\rm es}(t^{\prime})\beta_{\rm ej}/c)$.
Similarly, the velocity of the jet shock in $K^{\prime}$ is determined by the relative speed between the jet ($\beta_{\rm j}$) and the SN ejecta ($\beta_{\rm ej}$), expressing as $v^{\prime}_{\rm js}(t^{\prime})=\beta_{\rm rel}c$, where the relative velocity $\beta_{\rm rel}$ is given by $\beta_{\rm rel} = (\beta_{\rm j}-\beta_{\rm ej})/(1-\beta_{\rm j}\beta_{\rm ej})$.
\at{Note that for the ejecta emission phase, frame transformations are negligible since the ejecta has not yet been significantly accelerated.}

In $K^{\prime}$, the energetic electrons may also escape from the acceleration region via diffusion and advection. 
The effective escape timescale is determined by the faster of these two processes, 
\begin{equation}
\adm{t_{\rm esc}^{\prime}(E^{\prime}, B^{\prime}_{\rm s}) = {\rm min} [t^{\prime}_{\rm diff}(E^{\prime}, B^{\prime}_{\rm s}), t^{\prime}_{\rm adv}], } \label{eq: t_esc} 
\end{equation} 
where the diffusion timescale is given by \adt{$t^{\prime}_{\rm diff}\approx {R_{\rm ej}^{\prime}}^{2}/D^{\prime}_{\rm s}$ \citep{Aharonian2004vhec.book.....A}}. In the flow frame, \remo{perpendicular sizes do not change while }parallel sizes transform as \am{$R_{\rm ej}^{\prime}= \Gamma_{\rm ej}R_{\rm ej}$}\remo{$R_{\rm ej}^{\prime}= R_{\rm ej}/\Gamma_{\rm ej}$}, we take $R_{\rm ej}^{\prime}$ as the characteristic size of the SN ejecta in $K^{\prime}$. 
The diffusion coefficient in the acceleration region is $D^{\prime}_{\rm s}(E^{\prime}, B^{\prime}_{\rm s})=cE^{\prime}/(3eB^{\prime}_{\rm s})$. 
We adopt the whole size of the ejecta $R_{\rm ej}$ as the scale of the acceleration region. 
The advection timescale accounts for electrons carried away by the shocked flow and is estimated as \am{$t^{\prime}_{\rm adv} \sim R^{\prime}_{\rm ej} / v^{\prime}_{\rm flow}$}\remo{$t^{\prime}_{\rm adv, //} \sim R^{\prime}_{\rm ej} / v^{\prime}_{\rm flow}$} for the direction parallel to the shock motion (accounting for Lorentz contraction)\remo{ and $t^{\prime}_{\rm adv, \perp} \sim R_{\rm ej} / v_{\rm flow}$ for the perpendicular direction}, where $v^{\prime}_{\rm flow} \approx c/3$ is the escape velocity. 
\remo{The effective advection timescale is determined by the shorter of the two processes, namely $t^{\prime}_{\rm adv} = t^{\prime}_{\rm adv, //}$.}
By balancing the acceleration, cooling, and escape timescales ${t^{\prime}_{\rm acc}}^{-1} \gtrsim {t^{\prime}_{\rm cool}}^{-1} + {t^{\prime}_{\rm esc}}^{-1}$, the maximum energy of the high-energy electrons is 
\begin{align}
E^{\prime}_{\rm e, max}(t^{\prime}) \lesssim 1.21\times {\left(\frac{v^{\prime}(t^{\prime})}{0.3c}\right)} \left(\frac{B^{\prime}_{\rm s}}{10\mu \rm G}\right)^{-\frac{1}{2}} \rm PeV.
\label{eq: Emax}
\end{align}
The resultant maximum synchrotron photon energy that the electrons may therefore radiate is
\begin{align}
E_{\rm \gamma, max}^{\prime}(t^{\prime}) \lesssim 41.12\times \left(\frac{B^{\prime}_{\rm s}}{1\mu \rm G}\right) {\left(\frac{E^{\prime}_{\rm e, max}}{1 \rm PeV}\right)}^{2} \rm keV. 
\label{eq: gamma_max}
\end{align}
This indicates that electrons, energized via the ejecta shock or jet shock can produce \xray emission in the magnetic fields of $\sim$ $\mu \rm G$. 
\adt{In $K$, the corresponding maximum energy for electrons and emitted photons are approximated as $E_{\rm e,max} \approx \Gamma_{\rm ej} E'_{\rm e,max}$ and $E_{\rm \gamma, max} \approx \delta_{\rm ej} E'_{\rm \gamma,max} / (1+z)$, where $\delta_{\rm ej} = [\Gamma_{\rm ej}(1-\beta_{\rm ej} \rm cos\theta_{\rm j})]^{-1}$.}

\adt{
We assume an isotropic electron energy distribution in $K^{\prime}$. 
The synchrotron and IC radiation luminosities are calculated in $K^{\prime}$. 
We derive the comoving luminosity $L^{\prime}_{\rm \nu^\prime}$ to the observed flux $F_{\rm \nu}$. 
The observed flux density at the frequency $\nu=\delta_{\rm ej}\nu^{\prime}/(1+z)$ is given by \citep{Begelman1984RvMP, Dermer1995ApJ}: }
\begin{equation} 
\adm{F_{\rm \nu}(\nu) = \frac{\delta_{\rm ej}^3 (1 + z)}{4\pi D_{\rm L}^2} L^{\prime}_{\rm \nu}(\nu^{\prime}),} \label{eq:flux_transform} 
\end{equation} 
\adt{where $D_{\rm L}$ is the luminosity distance.}

\section{\adt{Application to knot A in M87 jet}\re{Radiation from accelerated electrons in M 87 jet}}\label{section: application}

\subsection{\ad{Physical properties of knot A}\re{Application to knot A in M87 jet}}\label{subs: data analysis}
\adt{To explore the origin of extended X-ray emission,} we test the hypothesis that the jet-ejecta interaction produces the \xray emission in the knots of the RG jets. 
\adt{We perform this test using two critical diagnostics, particle acceleration efficiency and spatial extent, as the observed extended X-ray radiation from RG knots requires the model to account for the spatial scale of the radiation region.}
\re{This test is performed through two critical diagnostics: particle acceleration efficiency and spatial extent. }
The nearby radio galaxy M 87 provides an ideal laboratory for this study due to its proximity and high surface brightness across radio-to-X-ray wavelengths. We select its brightest feature, knot A, as a test case.

M 87 (Virgo A, NGC 4486, 3C 274) is a massive elliptical galaxy ($z=0.004277$) located at a luminosity distance $D_{\rm L}$ of 16.7 ± 0.6 $\rm Mpc$ (1$\arcsec$ = 78 pc) \citep{Jord2005ApJ, Blakeslee2009ApJ}. It hosts a jet with an estimated power of $L_{\rm j} \sim 10^{43}$-$10^{45}$ $\rm erg\ s^{-1}$ \citep{Reynolds1996MNRAS, Gasperin2012A&A, Mo2016A&A, Levinson2017MNRAS} and \adt{$\theta_{\rm j} = 20 \deg$}\re{a viewing angle ($\theta_{\rm j}$) of 20$\deg$} \citep{Acciari2009Sci}. 
It is known to host a supermassive black hole (SMBH) of mass of $\sim (3.5-6.6) \times 10^{9} \msun$ \citep{Macchetto1997ApJ, Gebhardt2011ApJ, Walsh2013ApJ}. 
M 87 has been detected at radio, optical, and \xray wavelengths, and is a \ad{prime target for studying} the properties of relativistic jet \citep{Doeleman2012Sci, Hada2016ApJ, Mertens2016A&A, Britzen2017A&A, Sun2018}. 

We apply this jet-ejecta interaction framework to the bright knot A in the M87 jet. 
We first assess the viability of the ejecta shock as the primary electron accelerator for knot A, assuming a jet power of $L_{\rm j} = 10^{44}~\mathrm{erg~s^{-1}}$, a value typical for RG jets \citep{Perucho2019Galax}. 
The initial radius where the jet halts the upstream expansion of the ejecta is $R_{\rm ej,0} \approx 13$ pc. 
The ejecta shock traverses the unshocked material over a timescale of $t_{\rm max} \sim 1.5$ kyr, by which time the ejecta expands to a radius of merely $R_{\rm ej}\sim 30$ pc. 
\re{Upon acceleration to very high energies, electrons will begin to escape from the shock region. }
We expect electrons to escape the shock region once they are accelerated to very high energies. To quantify this and account for its impact,\re{To account for this,} we estimate the diffusion length of electrons with a characteristic energy of 1 PeV in a 10 $\mu$G magnetic field. 
The resulting diffusion scale is $\sim 6$ pc, implying that the non-thermal emission is confined strictly to the shock region. 
Consequently, the ejecta shock scenario fails to explain the observed spatial extent of knot A ($\sim 60$ pc; see Appendix \ref{appendix_data}). 

\adt{We instead focus on the jet shock scenario.}
\re{We therefore turn our attention to the jet shock as the dominant accelerator. }
According to the semi-analytical approach of \citet{Vieyro2019A&A}, the SN ejecta takes approximately $6000$ yr to expand to a radius of $60$ pc, assuming a predominantly adiabatic expansion driven by the internal pressure of the shocked ejecta. 
However, recent high-resolution relativistic hydrodynamical simulations by \citet{Longo2025A&A} indicate that comparable scales are reached on a shorter timescale of $3000$ to $4000$ yr. 
Adopting a reference epoch of $t_{\rm dyn} \approx 3000$ yr as the characteristic dynamical timescale of the jet-ejecta interaction, we find that the SN ejecta is fully shocked and accelerated to a bulk velocity of $\beta_{\rm ej}\approx 0.43$ ($\addm{\delta_{\rm ej}\approx1.53}$). 
The derived velocity at an evolutionary stage ($t_{\rm dyn} \approx 3000$ yr) is consistent with that of the numerical simulation results ($\sim0.25c$ – $0.50c$) \citep{Longo2025A&A}. 
The derived velocity is also compatible with that of the observed proper motion $< 0.7c$ of knot A \citep{Snios2019ApJ}.
\re{Doppler boosting be considered to account for the jet inclination angle with respect to the line of sight,}
\adt{Thus, the jet shock is identified as the primary site for particle acceleration.}

We adopt the approach of \citet{Vieyro2019A&A} to model the non-thermal emission from knot A, focusing on the jet shock as the dominant accelerator. 
We assume that a fraction of the kinetic power of the jet is dissipated at the jet shock and converted into accelerating a population of relativistic electrons. 
The energetic electrons can produce emissions via synchrotron and IC emissions within the shocked region. 
\re{We assume that the kinetic energy flux dissipated by the jet serves as a proxy for the energy flux available for accelerating non-thermal particles.}
In $K^{\prime}$, the jet shock luminosity represents the rate of kinetic energy dissipation at the shock front, and can be estimated as 
\begin{align}
L_{\rm s}^{\prime} = \pi {R^{\prime}_{\rm ej}}^{2} \beta_{\rm rel}\Gamma_{\rm rel}(\Gamma_{\rm rel}h_{\rm j}-1)\rho_{\rm j}c^{3}, \label{eq: shock luminosity}
\end{align} 
where $h_{\rm j}=1.1$ is the jet enthalpy, and $\rho_{\rm j} = 6\times10^{-30}\rm g\ cm^{-3}$ is the jet density \citep{Longo2025A&A}. 
\re{The relative velocity between the jet and SN ejecta is $\beta_{\rm rel} = (\beta_{\rm j}-\beta_{\rm ej})/(1-\beta_{\rm j}\beta_{\rm ej})$ in $c$ units. }
\adt{The derived luminosity of the jet shock is $L^{\prime}_{\rm s}=3.20\times10^{43}\ \rm erg\ s^{-1}$.}
The injected non-thermal power can be expressed in $K^{\prime}$ as
\begin{align}
L_{\rm NT}^{\prime} = \eta_{\rm NT}L_{\rm s}^{\prime}, \label{eq: injected power}
\end{align}
where $\eta_{\rm NT}$ is the fraction of dissipated kinetic energy converted into non-thermal electrons. 

We assume that the cooling timescale for the LEs is significantly longer than the dynamical timescale of the jet-ejecta interaction in the large-scale jets of RGs. 
This component accounts for the cumulative population of low-energy electrons in the ejecta region. 
The luminosity injected by the shock into the LE population is estimated by time-averaging the accumulated energy over the dynamical timescale. \at{We adopt a characteristic advection timescale of $t^{\prime}_{\rm adv}\sim1000$ yr, corresponding to an escape velocity of $v^{\prime}_{\rm flow}\sim c/3$. It should be mentioned that this is an order-of-magnitude estimate, as some of the LEs may already be present in the pre-shock jet flow. The injection luminosity is given by }\am{$L^{\prime}_{\rm LE}=W^{\prime}_{\rm e,LE}/t^{\prime}_{\rm adv}$}\remo{$L^{\prime}_{\rm LE}=W^{\prime}_{\rm e,LE}\Gamma_{\rm ej}/t_{\rm dyn}$}, where $W^{\prime}_{\rm e, LE}$ is the total energy of the LE population. 
The injected luminosity of the HE population is determined by integrating the differential injection rate, 
\begin{equation} 
\adm{L^{\prime}_{\rm HE} = \int^{E^{\prime}_{\rm e, max}}_{E^{\prime}_{\rm e, min}}Q^{\prime}_{\rm HE}(E^{\prime})E^{\prime}dE^{\prime}, }\label{eq: LE_luminosity}
\end{equation} 
\adt{where $E^{\prime}_{\rm e, min} = 1$ TeV is the minimum energy of the HE population.}
\adt{The total injected luminosity of the HEs is also given by}
\begin{equation} 
\adm{L^{\prime}_{\rm NT} = L^{\prime}_{\rm LE} + L^{\prime}_{\rm HE}.} 
\label{eq: total_injected_luminosity} 
\end{equation}

\subsection{SED fitting} \label{subs: fitting}
Adopting the electron distributions \adt{($N^{\prime}_{\rm LE}(E^{\prime})$ for the LE population and $N^{\prime}_{\rm HE}(E^{\prime})$ for the HE population)} described in Section \ref{acceleration}, we model the multiwavelength SED of knot A with two electron populations. 
\adt{
In $K^{\prime}$, we assume the two populations are co-spatial, sharing the same $B^{\prime}_{\rm s}$ and derived parameter $\eta_{\rm NT}$.
The two populations are characterized as follows: (1) The LE population, responsible for radio-to-optical emission, is described by $W^{\prime}_{\rm e, LE}$ and the spectral parameters ($\alpha_{\rm LE}$ and $E^{\prime}_{\rm cut, LE}$), these are treated as the free parameters in the fitting. 
This phenomenological approach provides the flexibility needed to obtain a good fit while avoiding biases that could arise from uncertainties in the cooling and escape timescales. 
(2) The HE population, responsible for the \xray emission, is fitted with two free parameters: the total steady-state energy $W^{\prime}_{\rm e, HE}$ of the HE population and the injection index $\alpha_{\rm HE}$. }
\re{The radio-to-optical emission is attributed to synchrotron radiation from a phenomenological LE population. 
The \xray data are explained by the synchrotron emission from a second population of accelerated electrons, whose properties are derived from our dynamical model (Eqs. \ref{eq: injected rate}–\ref{eq: steady state}). }
We perform the spectral fitting using the open-source package Naima \citep{Zabalza2015}, which provides numerical calculations of non-thermal emission from relativistic electron populations and performs Bayesian spectral fitting by Markov Chain Monte Carlo (MCMC) methods. 
The synchrotron and IC emission are calculated by Naima. 
\re{The total energy of the LE population $W_{\rm e, 1}$, the total energy of the HE population $W_{\rm e, 2}$, $\alpha_{1}$, $E^{\prime}_{\rm cut1}$, $\alpha_{2}$, $\zeta_{\rm ej}$, and $\eta_{\rm NT}$ are set as free parameters. }

\ad{Given the large projected distance ($\sim 1$ kpc) of knot A from the nucleus, the \gray emissions primarily originate from two processes: inverse Compton scattering of cosmic microwave background photons (IC/CMB) and the synchrotron self-Compton (SSC) process. 
We also account for the absorption caused by the extragalactic background light (EBL) \citep{Dom2011} to fit the \xray\ to \gray\ emissions. }
Nevertheless, we note that the \gray emission region of M 87 cannot be resolved, and the \gray emission may originate from the jet or the core. Thus, the \gray data is only treated as upper limits for the knot A in the modeling. 

\begin{figure}
  	\centering
    \includegraphics[width=1.0\linewidth]{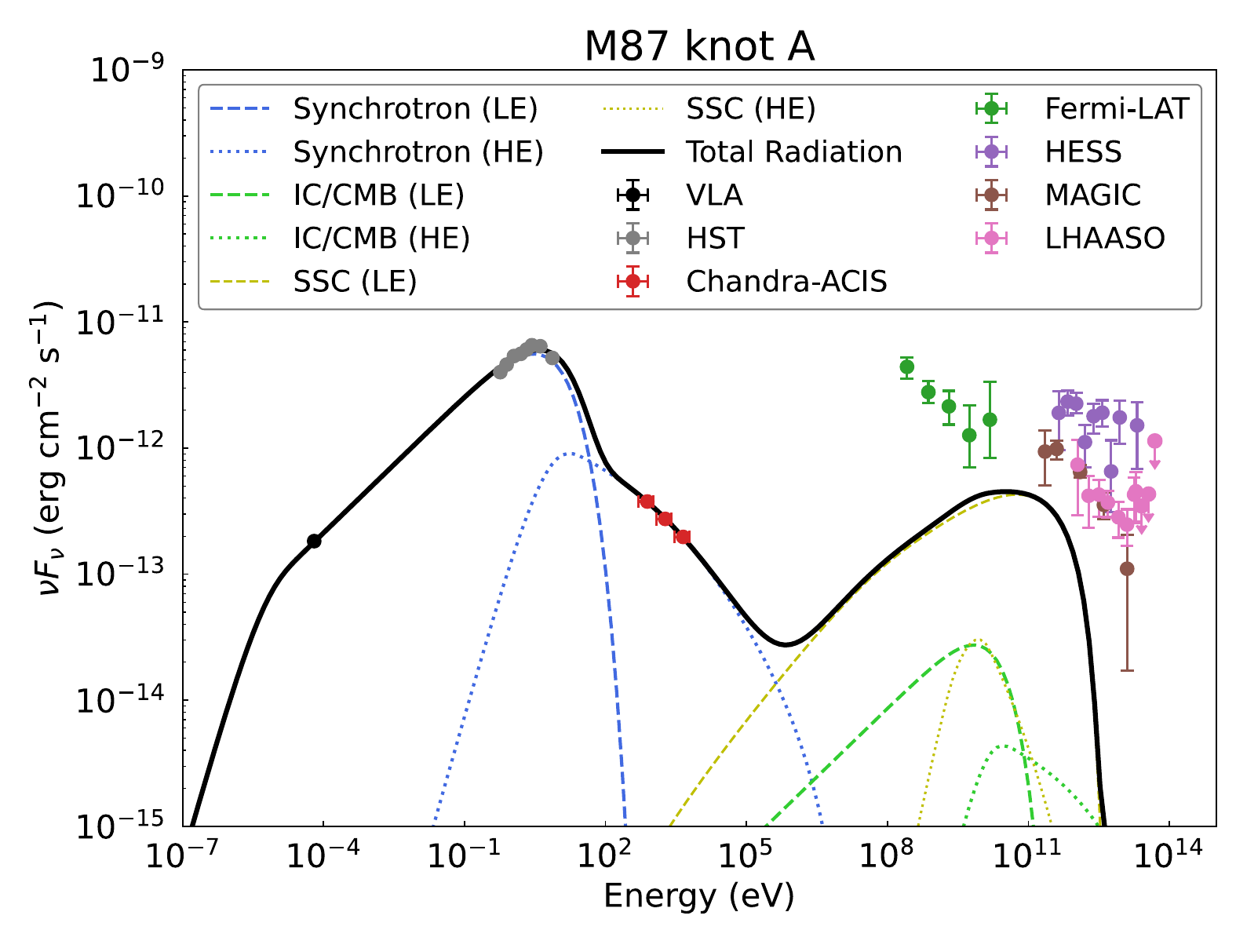}   
    \caption{The multiwavelength SED of knot A in M 87 reproduced by synchrotron and IC/CMB radiation of two electron populations in the jet-ejecta interaction framework. 
    The solid lines show the total non-thermal emission from two electron populations.
    The blue, green, and yellow dashed lines represent the synchrotron, IC/CMB, and SSC radiation of the LE population, respectively. 
    The blue, green, and yellow dotted lines are the synchrotron, IC/CMB, and SSC radiation of the HE population, respectively. 
    The red points are the data in the \xray energy band that we analyzed in this paper. 
    See the \addt{text} for details on the data from other instruments.
    } 
    \label{fig:SED} 
\end{figure}

The best-fitting SED of knot A is shown in Figure \ref{fig:SED}. 
The lines represent the SED fitting using the radiation model with the maximum-likelihood parameter values.
The individual contributions by the two electron populations are marked with dotted and dashed lines, respectively. 
The red points represent \chandra\ \xray\ data (0.5$-$7.0 keV) analyzed in \ad{Appendix \ref{appendix_data}.} 
Data from other instruments are marked as follows: black (Very Large Array, VLA) \citep{Perlman2001} and gray (Hubble Space Telescope, HST) \citep{Perlman2001}, green (\add{Fermi Large Area Telescope}, \fermi) \citep{Zhang2018a}, purple (H.E.S.S.), brown (Major Atmospheric Gamma-ray Imaging Cherenkov, MAGIC) \citep{MAGIC2020MNRAS}, and pink (Large High Altitude Air Shower Observatory, LHAASO) \citep{Cao2024ApJ}, respectively.
The corner plot for knot A (Figure \ref{fig:corner}) illustrates the parameter correlations and posterior distributions, showing that all free parameters are well-constrained and approximately follow Gaussian distributions.



\begin{table*}
\begin{center}
\caption{Best-fit parameters from the fitting of SED. }
\begin{threeparttable}

\renewcommand{\arraystretch}{1.6}
\begin{tabular}{c|ccccccc}
\hline\hline
\centering
\multirow{2}{*}{} & $W^{\prime}_{\rm e,LE}$ &$\alpha_{\rm LE}$& $E^{\prime}_{\rm cut, LE}$ &$W^{\prime}_{\rm e, HE}$&$\alpha_{\rm HE}$&$\zeta_{\rm ej}$\\
&[$\times 10^{53}\ \rm erg$]&  & [TeV] & [$\times\ 10^{51}\ \rm erg$]& & & \\\hline
M 87 - knot A & 3.51$^{+0.10}_{-0.08}$& 2.28$\pm$0.01 & 1.66$\pm$0.02 & $1.54^{+0.03}_{-0.06}$ &3.28$\pm$0.01& 0.26$\pm$0.01\\ 

\hline
\hline
\end{tabular}
\end{threeparttable}
 
\label{tab: free parameters}    
\end{center}

\end{table*}

\renewcommand{\arraystretch}{1.6}
\begin{table*}
\begin{center}
\caption{Derived dynamical parameters of the jet shock and accelerated electron.}
\begin{threeparttable}

\begin{tabular}{c|ccccccc}
\hline\hline
\centering
\multirow{2}{*}{} &$\eta_{\rm NT}$& $L^{\prime}_{\rm NT}$ & $L^{\prime}_{\rm LE}$ & $L^{\prime}_{\rm HE}$ &  $E^{\prime}_{\rm e, max}$& $B^{\prime}_{\rm s}$ \\
& &[$\times 10^{43}\ \rm erg\ s^{-1}$]&[$\times 10^{42}\ \rm erg\ s^{-1}$]&[$\times 10^{41}\ \rm erg\ s^{-1}$]& [PeV]&[$\mu \rm G$] \\
\hline
M 87 - knot A & 0.33$\pm$0.01& 1.07$\pm${0.01} & 9.75$\pm${0.01} & 9.02$\pm${0.01} & $1.05^{+0.17}_{-0.16}$&$72.3\pm0.01$\\ 
\hline
\hline
\end{tabular}
\end{threeparttable}
 
\label{tab: derived parameters}    
\end{center}

\end{table*}
\adt{The best-fitting values and derived dynamical parameters for knot A are summarized in Table \ref{tab: free parameters} and Table \ref{tab: derived parameters}, respectively.
The LE population exhibits a spectral index of $\alpha_{\rm LE} = 2.28 \pm 0.01$.
The derived cut-off energy $E^{\prime}_{\rm cut, LE} = 1.66\pm{0.02}$ TeV. 
The total energy is fitted with $W^{\prime}_{\rm e,LE}= 3.51^{+0.10}_{-0.08}\times10^{53}\ \rm erg$, corresponding to an injection luminosity of \at{$L^{\prime}_{\rm LE} = (9.75\pm0.01)\times10^{42}\ \rm erg\ s^{-1}$}\remo{$L^{\prime}_{\rm LE} = (3.25\pm0.01)\times10^{42}\ \rm erg\ s^{-1}$} that accounts for a power fraction of \am{$L^{\prime}_{\rm LE}/L^{\prime}_{\rm s}\approx30.6\%$}\remo{$L^{\prime}_{\rm LE}/L^{\prime}_{\rm s}\approx10.2\%$}. 
For the HE population, the best-fit spectral index is $\alpha_{\rm HE}=3.28\pm0.01$, the derived total energy and injection luminosity for this component are $W^{\prime}_{\rm e,HE}= 1.54\times10^{51}\ \rm erg$ and $L^{\prime}_{\rm HE} = (9.02\pm0.01)\times10^{41}\ \rm erg\ s^{-1}$, representing $L^{\prime}_{\rm HE}/L^{\prime}_{\rm s}\approx2.8\%$ of the shock luminosity.
The fraction of dissipated kinetic energy converted into non-thermal electrons is derived to be \am{$\eta_{\rm NT} = 0.33 \pm0.01$}\remo{$\eta_{\rm NT} = 0.13 \pm0.01$}. 
The magnetic field in the emitting region is $B^{\prime}_{\rm s} = 72.3 \pm0.01\ \mu \rm G$, 1 order of magnitude lower than the equipartition magnetic field range of $250 - 320\ \mu \rm G$ estimated for the emitter rest frame \citep{Marshall2002ApJ}. 
The injected electrons can be energized by the jet shock up to $1.05^{+0.17}_{-0.16}$ PeV. }

\re{The best-fitting values and uncertainties of model parameters are shown in Table \ref{tab: free parameters}. 
The radio-to-optical SED of knot A originates from synchrotron radiation of a LE population, exhibiting a spectral index $\alpha_{1} = 2.28 \pm 0.01$ and a cut-off energy $E_{\rm cut1} = 1.96\pm{0.03}$ TeV. 
The \xray SED is fitted with the spectral index $\alpha_{2}=3.52\pm0.01$, $\zeta_{\rm ej} = 0.11\pm0.01$, and $\eta_{\rm NT} = 0.09 \pm0.01$.
The derived dynamical parameters of jet and SN ejecta are shown in Table \ref{tab: derived parameters}. 
At $t_{\rm dyn} \sim 3000$ yr, the injected power of non-thermal electrons is calculated to be $L^{\prime}_{\rm NT}=(3.19\pm{0.01})\times10^{42}\ \rm erg\ s^{-1}$.
The total energy of the HE population is derived to be $W^{\prime}_{\rm e,2}= 1.54\times10^{51}\ \rm erg\ s^{-1}$. 
The magnetic field in the shock region is $B^{\prime}_{\rm s} = 30.1 \pm0.01\ \mu \rm G$, 1 order of magnitude lower than the equipartition magnetic field range of $250 - 320\ \mu \rm G$ estimated for the emitter rest frame. 
The injected electrons can be energized by the jet shock to $1.28^{+0.25}_{-0.12}$ PeV.}   

\section{Summary and discussion} \label{section:conclusion}
\ad{In this work, we investigated the feasibility of the jet-ejecta interaction scenario as an explanation for X-ray knots, focusing on two critical factors: (1) particle acceleration efficiency, and (2) the spatial extent of the X-ray knot. 
Theoretical and numerical studies have established the dynamical evolution of SN explosions within RG jets \citep{Vieyro2019A&A, Bosch-Ramon2023A&A, Longo2025A&A}, supporting the physical plausibility of such violent interactions.
Such collisions drive a distinct interaction characterized by two shocks: an ejecta shock propagating into and heating the ejecta, and a jet shock decelerating the incoming jet flow.  
The ejecta shock acts as an efficient particle accelerator, particularly during the early phases of the interaction, while the jet shock persists as a dominant, long-lived accelerator throughout the entire duration of the jet–ejecta interaction.
} 

M87 is one of the closest RGs with well-resolved jet knots (e.g., knot A), making it an ideal target for testing the jet-ejecta interaction framework. 
Applying this framework to knot A in the M87 jet, we first explore the ejecta shock as a potential accelerator. 
Assuming a jet power of $10^{44} \rm \ erg\ s^{-1}$, the ejecta shock scenario \addt{is likely disfavored} for knot A because the ejecta itself expands to only $\sim 30$ pc at $t_{\rm max}\sim 1.5$ kyr, rendering it incapable of producing the observed $\sim 60$ pc \xray structure.
In contrast, the jet shock at $t_{\rm dyn} \sim 3000$ yr naturally accounts for the spatial extent of the \xray emission from knot A, with the interaction region having expanded to match the observed scale of the knot \citep{Longo2025A&A}. 
The bulk motion of the shocked jet-ejecta mixture is estimated to reach $\beta_{\rm ej} \sim 0.43$. 
We model the multiwavelength emission of knot A within the jet-shock acceleration scenario, providing a leptonic interpretation of the SED from radio to X-rays. 
Our model incorporates the Doppler boosting effects for a viewing angle of $20^\circ$.  
In the jet-shock acceleration model, the multiwavelength SED (Figure~\ref{fig:SED}) can be satisfactorily reproduced by synchrotron and IC radiation with typical jet parameters (Tables~\ref{tab: free parameters} and ~\ref{tab: derived parameters}) for knot A in the M 87 jet. 
The SED fitting reveals two critical properties: (1) the derived magnetic field is far below equipartition, and (2) injected electrons are accelerated to $\sim$ PeV energies by the jet shock, and their synchrotron radiation then produces the observed \xray emissions. 
The required injected non-thermal luminosity is significantly less than the available jet shock luminosity in \addt{$K^{\prime}$}. 
\adt{The derived magnetic field is significantly below the equipartition value, implying a low magnetic energy density that may be insufficient to support magnetic reconnection as the primary acceleration mechanism. }\re{These properties suggest that energy dissipation at the shock front primarily favors particle acceleration over magnetic field amplification.}


Although we test the model using the knot A in M 87, the jet-ejecta interaction scenario is likely applicable to the broader population of RGs. 
The spatial scale of such interactions is related to the jet dynamics: the expansion radius of the SN ejecta ($R_{\rm ej}$) is governed by the pressure equilibrium with the jet flow and is thus intrinsically dependent on the jet power ($L_{\rm j}$). 
The jet power of M 87 ($L_{\rm j}\sim10^{44} \rm erg\ s^{-1}$) is characteristic of the transition between Fanaroff-Riley class I and II sources \citep{Bosch-Ramon2023A&A}, suggesting that the spatial scales and energetics derived in our model \addt{may be} representative for \at{AGN}\remo{the relativistic} jets. 
The leptonic model successfully reproduces the multi-wavelength SED while requiring \remo{only }a \at{significant}\remo{moderate} fraction \am{$L^{\prime}_{\rm NT}/L^{\prime}_{\rm s} \sim 33.4\%$}\remo{$L^{\prime}_{\rm NT}/L^{\prime}_{\rm s} \sim 13\%$}, suggesting that the sufficient energy budget may potentially drive hadronic acceleration. 
\at{The derived non-thermal efficiency indicates that a significant fraction of the dissipated kinetic energy is transferred to the accelerated particles.}
\remo{This efficiency is compatible with the typical values of $10\%-20\%$ for ion acceleration in the kinetic simulations of diffusive shock acceleration \citep{Caprioli2014ApJ}.}
This is consistent with the scenario that jet-ejecta interactions convert a significant portion of the jet power into ultra-high-energy cosmic-ray (UHECRs) \citep{Bosch-Ramon2023A&A}.  
The \addt{dynamical evolution} of SN ejecta can transport baryon-rich ejecta material into the relativistic flow. 
Given the efficient electron acceleration inferred from the \xray modeling of M 87, \addt{the jet shock may} naturally accelerate heavier nuclei, thereby serving as a potential site for UHECR acceleration or, at the very least, providing seed particles for further acceleration elsewhere in the jet.
For jet-shock acceleration, protons and other nuclei can also be accelerated with a spectrum similar to that of electrons, while the escaping protons and other nuclei could have an even harder spectrum. 
\addt{Assuming that the acceleration of protons and nuclei proceeds in the same manner and with the same high efficiency as the high-energy electron acceleration, 
}the maximum energy is estimated to be $E^{\prime}_{\rm p, max} \sim 10Z\left(\frac{v^{\prime}(t^{\prime})}{c}\right)\left(\frac{R^{\prime}_{\rm ej}}{100\rm\ pc}\right)\left(\frac{B^{\prime}_{\rm s}}{0.3B_{\rm eq}}\right)\left(\frac{L_{\rm j}}{10^{44}\rm\ erg\ s^{-1}}\right)^{\frac{1}{2}}\left(\frac{R_{\rm j}}{100\rm\ pc}\right)^{-1} \rm EeV$ \citep{Bosch-Ramon2023A&A}, where $Z$ is the atomic number. 
This supports that the jet-ejecta interaction is a favorable mechanism for UHECR acceleration within the RG jets. 

\section*{acknowledgements}   \label{acknowledgements}
This work is supported by the Guangxi Science Foundation (grant No. 2024GXNSFBA010375), Guangxi Key Research and Development Program (Guike FN2504240040), 
the National Natural Science Foundation of China (NSFC, grant No. 12133003 and 12494573), and the Guangxi Talent Program (“Highland of Innovation Talents”), the Bagui Scholars Programme (W.X.-G., GXR-6BG2424001). 
\add{We thank the referee for valuable comments.}
We are grateful to J. Zhang for the constructive discussions.  
We thank M. Jiang for helpful discussions on the data. 


\section{data availability} \label{data availability}
The \chandra\ ACIS data used in work are publicly available, which is provided online by the Chandra Data Archive\footnote{\url{https://cda.harvard.edu/chaser/mainEntry.do}}. 

\bibliographystyle{mnras}
\bibliography{ms}

@ARTICLE{Wang2023MNRAS,
       author = {{Wang}, Jie-Shuang and {Reville}, Brian and {Mizuno}, Yosuke and {Rieger}, Frank M. and {Aharonian}, Felix A.},
        title = "{Particle acceleration in shearing flows: the self-generation of turbulent spine-sheath structures in relativistic magnetohydrodynamic jet simulations}",
      journal = {\mnras},
         year = 2023,
        month = feb,
       volume = {519},
       number = {2},
        pages = {1872-1880},
          doi = {10.1093/mnras/stac3616},
archivePrefix = {arXiv},
       eprint = {2212.03226},
 primaryClass = {astro-ph.HE},
       adsurl = {https://ui.adsabs.harvard.edu/abs/2023MNRAS.519.1872W}
}

@ARTICLE{Harris2006,
       author = {{Harris}, D.~E. and {Krawczynski}, Henric},
        title = "{X-Ray Emission from Extragalactic Jets}",
      journal = {\araa},
         year = 2006,
        month = sep,
       volume = {44},
       number = {1},
        pages = {463-506},
          doi = {10.1146/annurev.astro.44.051905.092446},
archivePrefix = {arXiv},
       eprint = {astro-ph/0607228},
 primaryClass = {astro-ph},
       adsurl = {https://ui.adsabs.harvard.edu/abs/2006ARA&A..44..463H}
}

@ARTICLE{Zhang2018b,
       author = {{Zhang}, Jin and {Zhang}, Hai-Ming and {Yao}, Su and {Guo}, Sheng-Chu and {Lu}, Rui-Jing and {Liang}, En-Wei},
        title = "{Jet Radiation Properties of 4C +49.22: from the Core to Large-scale Knots}",
      journal = {\apj},
         year = 2018,
        month = oct,
       volume = {865},
       number = {2},
          eid = {100},
        pages = {100},
          doi = {10.3847/1538-4357/aadd0b},
archivePrefix = {arXiv},
       eprint = {1808.07978},
 primaryClass = {astro-ph.HE},
       adsurl = {https://ui.adsabs.harvard.edu/abs/2018ApJ...865..100Z}
}

@ARTICLE{Zhang2018a,
       author = {{Zhang}, Jin and {Du}, Shen-shi and {Guo}, Sheng-Chu and {Zhang}, Hai-Ming and {Chen}, Liang and {Liang}, En-Wei and {Zhang}, Shuang-Nan},
        title = "{Examining the High-energy Radiation Mechanisms of Knots and Hotspots in Active Galactic Nucleus Jets}",
      journal = {\apj},
         year = 2018,
        month = may,
       volume = {858},
       number = {1},
          eid = {27},
        pages = {27},
          doi = {10.3847/1538-4357/aab9b2},
archivePrefix = {arXiv},
       eprint = {1803.08639},
 primaryClass = {astro-ph.HE},
       adsurl = {https://ui.adsabs.harvard.edu/abs/2018ApJ...858...27Z}
}

@misc{1,
   author = {A. Mossman},
   title = {XJET Home Page},
   howpublished = {\url{https://hea-www.harvard.edu/XJET}},
   year = 2001
}

@misc{2,
   author = {the Smithsonian Astrophysical Observatory},
   title = {Reprojecting Aspect - CIAO 4.14},
   howpublished = {\url{https://cxc.cfa.harvard.edu/ciao/threads/index.html}},
   year = 2003
}

@INPROCEEDINGS{Zabalza2015,
       author = {{Zabalza}, V.},
        title = "{Naima: a Python package for inference of particle distribution properties from nonthermal spectra}",
    booktitle = {34th International Cosmic Ray Conference (ICRC2015)},
         year = 2015,
       series = {International Cosmic Ray Conference},
       volume = {34},
        month = jul,
          eid = {922},
        pages = {922},
          doi = {10.22323/1.236.0922},
archivePrefix = {arXiv},
       eprint = {1509.03319},
 primaryClass = {astro-ph.HE},
       adsurl = {https://ui.adsabs.harvard.edu/abs/2015ICRC...34..922Z}
}

@ARTICLE{Wang2021MNRAS,
       author = {{Wang}, Jie-Shuang and {Reville}, Brian and {Liu}, Ruo-Yu and {Rieger}, Frank M. and {Aharonian}, Felix A.},
        title = "{Particle acceleration in shearing flows: the case for large-scale jets}",
      journal = {\mnras},
         year = 2021,
        month = jul,
       volume = {505},
       number = {1},
        pages = {1334-1341},
          doi = {10.1093/mnras/stab1458},
archivePrefix = {arXiv},
       eprint = {2105.08600},
 primaryClass = {astro-ph.HE},
       adsurl = {https://ui.adsabs.harvard.edu/abs/2021MNRAS.505.1334W}
}

@ARTICLE{Rieger2019,
       author = {{Rieger}, Frank M. and {Duffy}, Peter},
        title = "{Particle Acceleration in Shearing Flows: Efficiencies and Limits}",
      journal = {\apjl},
         year = 2019,
        month = dec,
       volume = {886},
       number = {2},
          eid = {L26},
        pages = {L26},
          doi = {10.3847/2041-8213/ab563f},
archivePrefix = {arXiv},
       eprint = {1911.05348},
 primaryClass = {astro-ph.HE},
       adsurl = {https://ui.adsabs.harvard.edu/abs/2019ApJ...886L..26R}
}

@ARTICLE{Zhang2009,
       author = {{Zhang}, Weiqun and {MacFadyen}, Andrew and {Wang}, Peng},
        title = "{Three-Dimensional Relativistic Magnetohydrodynamic Simulations of the Kelvin-Helmholtz Instability: Magnetic Field Amplification by a Turbulent Dynamo}",
      journal = {\apjl},
         year = 2009,
        month = feb,
       volume = {692},
       number = {1},
        pages = {L40-L44},
          doi = {10.1088/0004-637X/692/1/L40},
archivePrefix = {arXiv},
       eprint = {0811.3638},
 primaryClass = {astro-ph},
       adsurl = {https://ui.adsabs.harvard.edu/abs/2009ApJ...692L..40Z}
}

@ARTICLE{Georganopoulos2016,
       author = {{Georganopoulos}, Markos and {Meyer}, Eileen and {Perlman}, Eric},
        title = "{Recent Progress in Understanding the Large Scale Jets of Powerful Quasars}",
      journal = {Galaxies},
         year = 2016,
        month = nov,
       volume = {4},
       number = {4},
        pages = {65},
          doi = {10.3390/galaxies4040065},
       adsurl = {https://ui.adsabs.harvard.edu/abs/2016Galax...4...65G}
}

@ARTICLE{Sun2018,
       author = {{Sun}, Xiao-Na and {Yang}, Rui-Zhi and {Rieger}, Frank M. and {Liu}, Ruo-Yu and {Aharonian}, Felix},
        title = "{Energy distribution of relativistic electrons in the kiloparsec scale jet of M 87 with Chandra}",
      journal = {\aap},
         year = 2018,
        month = may,
       volume = {612},
          eid = {A106},
        pages = {A106},
          doi = {10.1051/0004-6361/201731716},
archivePrefix = {arXiv},
       eprint = {1712.06390},
 primaryClass = {astro-ph.HE},
       adsurl = {https://ui.adsabs.harvard.edu/abs/2018A&A...612A.106S}
}

@ARTICLE{Jester2007,
       author = {{Jester}, Sebastian and {Meisenheimer}, Klaus and {Martel}, Andr{\'e} R. and {Perlman}, Eric S. and {Sparks}, William B.},
        title = "{Hubble Space Telescope far-ultraviolet imaging of the jet in 3C273: a common emission component from optical to X-rays}",
      journal = {\mnras},
         year = 2007,
        month = sep,
       volume = {380},
       number = {2},
        pages = {828-834},
          doi = {10.1111/j.1365-2966.2007.12120.x},
archivePrefix = {arXiv},
       eprint = {0706.2564},
 primaryClass = {astro-ph},
       adsurl = {https://ui.adsabs.harvard.edu/abs/2007MNRAS.380..828J}
}

@ARTICLE{Weisskopf2002,
       author = {{Weisskopf}, M.~C. and {Brinkman}, B. and {Canizares}, C. and {Garmire}, G. and {Murray}, S. and {Van Speybroeck}, L.~P.},
        title = "{An Overview of the Performance and Scientific Results from the Chandra X-Ray Observatory}",
      journal = {\pasp},
         year = 2002,
        month = jan,
       volume = {114},
       number = {791},
        pages = {1-24},
          doi = {10.1086/338108},
archivePrefix = {arXiv},
       eprint = {astro-ph/0110308},
 primaryClass = {astro-ph},
       adsurl = {https://ui.adsabs.harvard.edu/abs/2002PASP..114....1W}
}

@ARTICLE{Perlman2001,
       author = {{Perlman}, Eric S. and {Biretta}, John A. and {Sparks}, William B. and {Macchetto}, F. Duccio and {Leahy}, J. Patrick},
        title = "{The Optical-Near-Infrared Spectrum of the M87 Jet fromHubble Space Telescope Observations}",
      journal = {\apj},
         year = 2001,
        month = apr,
       volume = {551},
       number = {1},
        pages = {206-222},
          doi = {10.1086/320052},
archivePrefix = {arXiv},
       eprint = {astro-ph/0012044},
 primaryClass = {astro-ph},
       adsurl = {https://ui.adsabs.harvard.edu/abs/2001ApJ...551..206P}
}

@ARTICLE{Dom2011,
       author = {{Dom{\'\i}nguez}, A. and {Primack}, J.~R. and {Rosario}, D.~J. and {Prada}, F. and {Gilmore}, R.~C. and {Faber}, S.~M. and {Koo}, D.~C. and {Somerville}, R.~S. and {P{\'e}rez-Torres}, M.~A. and {P{\'e}rez-Gonz{\'a}lez}, P. and {Huang}, J. -S. and {Davis}, M. and {Guhathakurta}, P. and {Barmby}, P. and {Conselice}, C.~J. and {Lozano}, M. and {Newman}, J.~A. and {Cooper}, M.~C.},
        title = "{Extragalactic background light inferred from AEGIS galaxy-SED-type fractions}",
      journal = {\mnras},
         year = 2011,
        month = feb,
       volume = {410},
       number = {4},
        pages = {2556-2578},
          doi = {10.1111/j.1365-2966.2010.17631.x},
archivePrefix = {arXiv},
       eprint = {1007.1459},
 primaryClass = {astro-ph.CO},
       adsurl = {https://ui.adsabs.harvard.edu/abs/2011MNRAS.410.2556D}
}

@ARTICLE{Jester2006,
       author = {{Jester}, Sebastian and {Harris}, D.~E. and {Marshall}, Herman L. and {Meisenheimer}, Klaus},
        title = "{New Chandra Observations of the Jet in 3C 273. I. Softer X-Ray than Radio Spectra and the X-Ray Emission Mechanism}",
      journal = {\apj},
         year = 2006,
        month = sep,
       volume = {648},
       number = {2},
        pages = {900-909},
          doi = {10.1086/505962},
archivePrefix = {arXiv},
       eprint = {astro-ph/0605529},
 primaryClass = {astro-ph},
       adsurl = {https://ui.adsabs.harvard.edu/abs/2006ApJ...648..900J}
}

@ARTICLE{HESS20,
       author = {{H.~E.~S.~S. Collaboration} and {Abdalla}, H. and {Adam}, R. and {Aharonian}, F. and {Ait Benkhali}, F. and {Ang{\"u}ner}, E.~O. and {Arakawa}, M. and {Arcaro}, C. and {Armand}, C. and {Ashkar}, H. and {Backes}, M. and {Barbosa Martins}, V. and {Barnard}, M. and {Becherini}, Y. and {Berge}, D. and {Bernl{\"o}hr}, K. and {Blackwell}, R. and {B{\"o}ttcher}, M. and {Boisson}, C. and {Bolmont}, J. and {Bonnefoy}, S. and {Bregeon}, J. and {Breuhaus}, M. and {Brun}, F. and {Brun}, P. and {Bryan}, M. and {B{\"u}chele}, M. and {Bulik}, T. and {Bylund}, T. and {Capasso}, M. and {Caroff}, S. and {Carosi}, A. and {Casanova}, S. and {Cerruti}, M. and {Chand}, T. and {Chandra}, S. and {Chen}, A. and {Colafrancesco}, S. and {Cury{\l}o}, M. and {Davids}, I.~D. and {Deil}, C. and {Devin}, J. and {deWilt}, P. and {Dirson}, L. and {Djannati-Ata{\"\i}}, A. and {Dmytriiev}, A. and {Donath}, A. and {Doroshenko}, V. and {Drury}, L. O'C. and {Dyks}, J. and {Egberts}, K. and {Emery}, G. and {Ernenwein}, J. -P. and {Eschbach}, S. and {Feijen}, K. and {Fegan}, S. and {Fiasson}, A. and {Fontaine}, G. and {Funk}, S. and {F{\"u}{\ss}ling}, M. and {Gabici}, S. and {Gallant}, Y.~A. and {Gat{\'e}}, F. and {Giavitto}, G. and {Glawion}, D. and {Glicenstein}, J.~F. and {Gottschall}, D. and {Grondin}, M. -H. and {Hahn}, J. and {Haupt}, M. and {Heinzelmann}, G. and {Henri}, G. and {Hermann}, G. and {Hinton}, J.~A. and {Hofmann}, W. and {Hoischen}, C. and {Holch}, T.~L. and {Holler}, M. and {Horns}, D. and {Huber}, D. and {Iwasaki}, H. and {Jamrozy}, M. and {Jankowsky}, D. and {Jankowsky}, F. and {Jardin-Blicq}, A. and {Jung-Richardt}, I. and {Kastendieck}, M.~A. and {Katarzy{\'n}ski}, K. and {Katsuragawa}, M. and {Katz}, U. and {Khangulyan}, D. and {Kh{\'e}lifi}, B. and {King}, J. and {Klepser}, S. and {Klu{\'z}niak}, W. and {Komin}, N. and {Kosack}, K. and {Kostunin}, D. and {Kraus}, M. and {Lamanna}, G. and {Lau}, J. and {Lemi{\`e}re}, A. and {Lemoine-Goumard}, M. and {Lenain}, J. -P. and {Leser}, E. and {Levy}, C. and {Lohse}, T. and {Lypova}, I. and {Mackey}, J. and {Majumdar}, J. and {Malyshev}, D. and {Marandon}, V. and {Marcowith}, A. and {Mares}, A. and {Mariaud}, C. and {Mart{\'\i}-Devesa}, G. and {Marx}, R. and {Maurin}, G. and {Meintjes}, P.~J. and {Mitchell}, A.~M.~W. and {Moderski}, R. and {Mohamed}, M. and {Mohrmann}, L. and {Moore}, C. and {Moulin}, E. and {Muller}, J. and {Murach}, T. and {Nakashima}, S. and {de Naurois}, M. and {Ndiyavala}, H. and {Niederwanger}, F. and {Niemiec}, J. and {Oakes}, L. and {O'Brien}, P. and {Odaka}, H. and {Ohm}, S. and {de Ona Wilhelmi}, E. and {Ostrowski}, M. and {Oya}, I. and {Panter}, M. and {Parsons}, R.~D. and {Perennes}, C. and {Petrucci}, P. -O. and {Peyaud}, B. and {Piel}, Q. and {Pita}, S. and {Poireau}, V. and {Priyana Noel}, A. and {Prokhorov}, D.~A. and {Prokoph}, H. and {P{\"u}hlhofer}, G. and {Punch}, M. and {Quirrenbach}, A. and {Raab}, S. and {Rauth}, R. and {Reimer}, A. and {Reimer}, O. and {Remy}, Q. and {Renaud}, M. and {Rieger}, F. and {Rinchiuso}, L. and {Romoli}, C. and {Rowell}, G. and {Rudak}, B. and {Ruiz-Velasco}, E. and {Sahakian}, V. and {Saito}, S. and {Sanchez}, D.~A. and {Santangelo}, A. and {Sasaki}, M. and {Schlickeiser}, R. and {Sch{\"u}ssler}, F. and {Schulz}, A. and {Schutte}, H.~M. and {Schwanke}, U. and {Schwemmer}, S. and {Seglar-Arroyo}, M. and {Senniappan}, M. and {Seyffert}, A.~S. and {Shafi}, N. and {Shiningayamwe}, K. and {Simoni}, R. and {Sinha}, A. and {Sol}, H. and {Specovius}, A. and {Spir-Jacob}, M. and {Stawarz}, {\L}. and {Steenkamp}, R. and {Stegmann}, C. and {Steppa}, C. and {Takahashi}, T. and {Tavernier}, T. and {Taylor}, A.~M. and {Terrier}, R. and {Tiziani}, D. and {Tluczykont}, M. and {Trichard}, C. and {Tsirou}, M. and {Tsuji}, N. and {Tuffs}, R. and {Uchiyama}, Y. and {van der Walt}, D.~J. and {van Eldik}, C. and {van Rensburg}, C. and {van Soelen}, B. and {Vasileiadis}, G. and {Veh}, J. and {Venter}, C. and {Vincent}, P. and {Vink}, J. and {Voisin}, F. and {V{\"o}lk}, H.~J. and {Vuillaume}, T. and {Wadiasingh}, Z. and {Wagner}, S.~J. and {White}, R. and {Wierzcholska}, A. and {Yang}, R. and {Yoneda}, H. and {Zacharias}, M. and {Zanin}, R. and {Zdziarski}, A.~A. and {Zech}, A. and {Ziegler}, A. and {Zorn}, J. and {{\.Z}ywucka}, N.},
        title = "{Resolving acceleration to very high energies along the jet of Centaurus A}",
      journal = {\nat},
         year = 2020,
        month = jun,
       volume = {582},
       number = {7812},
        pages = {356-359},
          doi = {10.1038/s41586-020-2354-1},
archivePrefix = {arXiv},
       eprint = {2007.04823},
 primaryClass = {astro-ph.HE},
       adsurl = {https://ui.adsabs.harvard.edu/abs/2020Natur.582..356H}
}

@ARTICLE{Rieger2022a,
       author = {{Rieger}, Frank M. and {Duffy}, Peter},
        title = "{Particle Acceleration in Relativistic Shearing Flows: Energy Spectrum}",
      journal = {\apj},
         year = 2022,
        month = jul,
       volume = {933},
       number = {2},
          eid = {149},
        pages = {149},
          doi = {10.3847/1538-4357/ac729c},
archivePrefix = {arXiv},
       eprint = {2206.13098},
 primaryClass = {astro-ph.HE},
       adsurl = {https://ui.adsabs.harvard.edu/abs/2022ApJ...933..149R}
}

@ARTICLE{Breiding2017,
       author = {{Breiding}, Peter and {Meyer}, Eileen T. and {Georganopoulos}, Markos and {Keenan}, M.~E. and {DeNigris}, N.~S. and {Hewitt}, Jennifer},
        title = "{Fermi Non-detections of Four X-Ray Jet Sources and Implications for the IC/CMB Mechanism}",
      journal = {\apj},
         year = 2017,
        month = nov,
       volume = {849},
       number = {2},
          eid = {95},
        pages = {95},
          doi = {10.3847/1538-4357/aa907a},
archivePrefix = {arXiv},
       eprint = {1710.04250},
 primaryClass = {astro-ph.GA},
       adsurl = {https://ui.adsabs.harvard.edu/abs/2017ApJ...849...95B}
}

@ARTICLE{He2023MNRAS,
       author = {{He}, Jia-Chun and {Sun}, Xiao-Na and {Wang}, Jie-Shuang and {Rieger}, Frank M. and {Liu}, Ruo-Yu and {Liang}, En-Wei},
        title = "{Studying X-ray spectra from large-scale jets of FR II radio galaxies: application of shear particle acceleration}",
      journal = {\mnras},
         year = 2023,
        month = nov,
       volume = {525},
       number = {4},
        pages = {5298-5310},
          doi = {10.1093/mnras/stad2542},
archivePrefix = {arXiv},
       eprint = {2308.11370},
 primaryClass = {astro-ph.HE},
       adsurl = {https://ui.adsabs.harvard.edu/abs/2023MNRAS.525.5298H}
}

@ARTICLE{Vieyro2019A&A,
       author = {{Vieyro}, F.~L. and {Bosch-Ramon}, V. and {Torres-Alb{\`a}}, N.},
        title = "{Non-thermal emission resulting from a supernova explosion inside an extragalactic jet}",
      journal = {\aap},
         year = 2019,
        month = feb,
       volume = {622},
          eid = {A175},
        pages = {A175},
          doi = {10.1051/0004-6361/201833319},
archivePrefix = {arXiv},
       eprint = {1901.09003},
 primaryClass = {astro-ph.HE},
       adsurl = {https://ui.adsabs.harvard.edu/abs/2019A&A...622A.175V}
}

@ARTICLE{Blandford1979ApL,
       author = {{Blandford}, R.~D. and {Koenigl}, A.},
        title = "{A Model for the Knots in the M87 Jet}",
      journal = {\aplett},
         year = 1979,
        month = jan,
       volume = {20},
        pages = {15},
       adsurl = {https://ui.adsabs.harvard.edu/abs/1979ApL....20...15B}
}

@ARTICLE{Barkov2012ApJ,
       author = {{Barkov}, M.~V. and {Aharonian}, F.~A. and {Bogovalov}, S.~V. and {Kelner}, S.~R. and {Khangulyan}, D.},
        title = "{Rapid TeV Variability in Blazars as a Result of Jet-Star Interaction}",
      journal = {\apj},
         year = 2012,
        month = apr,
       volume = {749},
       number = {2},
          eid = {119},
        pages = {119},
          doi = {10.1088/0004-637X/749/2/119},
archivePrefix = {arXiv},
       eprint = {1012.1787},
 primaryClass = {astro-ph.HE},
       adsurl = {https://ui.adsabs.harvard.edu/abs/2012ApJ...749..119B}
}

@ARTICLE{Bosch-Ramon2015A&A,
       author = {{Bosch-Ramon}, V.},
        title = "{Non-thermal emission from standing relativistic shocks: an application to red giant winds interacting with AGN jets}",
      journal = {\aap},
         year = 2015,
        month = mar,
       volume = {575},
          eid = {A109},
        pages = {A109},
          doi = {10.1051/0004-6361/201425208},
archivePrefix = {arXiv},
       eprint = {1501.03118},
 primaryClass = {astro-ph.HE},
       adsurl = {https://ui.adsabs.harvard.edu/abs/2015A&A...575A.109B}
}

@ARTICLE{Bosch-Ramon2023A&A,
       author = {{Bosch-Ramon}, V.},
        title = "{The role of supernovae inside AGN jets in UHECR acceleration}",
      journal = {\aap},
         year = 2023,
        month = sep,
       volume = {677},
          eid = {L14},
        pages = {L14},
          doi = {10.1051/0004-6361/202347554},
archivePrefix = {arXiv},
       eprint = {2309.02239},
 primaryClass = {astro-ph.HE},
       adsurl = {https://ui.adsabs.harvard.edu/abs/2023A&A...677L..14B}
}

@ARTICLE{Acciari2009Sci,
       author = {{Acciari}, V.~A. and {Aliu}, E. and {Arlen}, T. and {Bautista}, M. and {Beilicke}, M. and {Benbow}, W. and {Bradbury}, S.~M. and {Buckley}, J.~H. and {Bugaev}, V. and {Butt}, Y. and {Byrum}, K. and {Cannon}, A. and {Celik}, O. and {Cesarini}, A. and {Chow}, Y.~C. and {Ciupik}, L. and {Cogan}, P. and {Cui}, W. and {Dickherber}, R. and {Fegan}, S.~J. and {Finley}, J.~P. and {Fortin}, P. and {Fortson}, L. and {Furniss}, A. and {Gall}, D. and {Gillanders}, G.~H. and {Grube}, J. and {Guenette}, R. and {Gyuk}, G. and {Hanna}, D. and {Holder}, J. and {Horan}, D. and {Hui}, C.~M. and {Humensky}, T.~B. and {Imran}, A. and {Kaaret}, P. and {Karlsson}, N. and {Kieda}, D. and {Kildea}, J. and {Konopelko}, A. and {Krawczynski}, H. and {Krennrich}, F. and {Lang}, M.~J. and {LeBohec}, S. and {Maier}, G. and {McCann}, A. and {McCutcheon}, M. and {Millis}, J. and {Moriarty}, P. and {Ong}, R.~A. and {Otte}, A.~N. and {Pandel}, D. and {Perkins}, J.~S. and {Petry}, D. and {Pohl}, M. and {Quinn}, J. and {Ragan}, K. and {Reyes}, L.~C. and {Reynolds}, P.~T. and {Roache}, E. and {Roache}, E. and {Rose}, H.~J. and {Schroedter}, M. and {Sembroski}, G.~H. and {Smith}, A.~W. and {Swordy}, S.~P. and {Theiling}, M. and {Toner}, J.~A. and {Varlotta}, A. and {Vincent}, S. and {Wakely}, S.~P. and {Ward}, J.~E. and {Weekes}, T.~C. and {Weinstein}, A. and {Williams}, D.~A. and {Wissel}, S. and {Wood}, M. and {Walker}, R.~C. and {Davies}, F. and {Hardee}, P.~E. and {Junor}, W. and {Ly}, C. and {Aharonian}, F. and {Akhperjanian}, A.~G. and {Anton}, G. and {Barres de Almeida}, U. and {Bazer-Bachi}, A.~R. and {Becherini}, Y. and {Behera}, B. and {Bernl{\"o}hr}, K. and {Bochow}, A. and {Boisson}, C. and {Bolmont}, J. and {Borrel}, V. and {Brucker}, J. and {Brun}, F. and {Brun}, P. and {B{\"u}hler}, R. and {Bulik}, T. and {B{\"u}sching}, I. and {Boutelier}, T. and {Chadwick}, P.~M. and {Charbonnier}, A. and {Chaves}, R.~C.~G. and {Cheesebrough}, A. and {Chounet}, L. -M. and {Clapson}, A.~C. and {Coignet}, G. and {Dalton}, M. and {Daniel}, M.~K. and {Davids}, I.~D. and {Degrange}, B. and {Deil}, C. and {Dickinson}, H.~J. and {Djannati-Ata{\"\i}}, A. and {Domainko}, W. and {Drury}, L.~O. 'C. and {Dubois}, F. and {Dubus}, G. and {Dyks}, J. and {Dyrda}, M. and {Egberts}, K. and {Emmanoulopoulos}, D. and {Espigat}, P. and {Farnier}, C. and {Feinstein}, F. and {Fiasson}, A. and {F{\"o}rster}, A. and {Fontaine}, G. and {F{\"u}{\ss}ling}, M. and {Gabici}, S. and {Gallant}, Y.~A. and {G{\'e}rard}, L. and {Gerbig}, D. and {Giebels}, B. and {Glicenstein}, J.~F. and {Gl{\"u}ck}, B. and {Goret}, P. and {G{\"o}hring}, D. and {Hauser}, D. and {Hauser}, M. and {Heinz}, S. and {Heinzelmann}, G. and {Henri}, G. and {Hermann}, G. and {Hinton}, J.~A. and {Hoffmann}, A. and {Hofmann}, W. and {Holleran}, M. and {Hoppe}, S. and {Horns}, D. and {Jacholkowska}, A. and {de Jager}, O.~C. and {Jahn}, C. and {Jung}, I. and {Katarzy{\'n}ski}, K. and {Katz}, U. and {Kaufmann}, S. and {Kendziorra}, E. and {Kerschhaggl}, M. and {Khangulyan}, D. and {Kh{\'e}lifi}, B. and {Keogh}, D. and {Klu{\'z}niak}, W. and {Kneiske}, T. and {Komin}, Nu. and {Kosack}, K. and {Lamanna}, G. and {Lenain}, J. -P. and {Lohse}, T. and {Marandon}, V. and {Martin}, J.~M. and {Martineau-Huynh}, O. and {Marcowith}, A. and {Maurin}, D. and {McComb}, T.~J.~L. and {Medina}, M.~C. and {Moderski}, R. and {Moulin}, E. and {Naumann-Godo}, M. and {de Naurois}, M. and {Nedbal}, D. and {Nekrassov}, D. and {Nicholas}, B. and {Niemiec}, J. and {Nolan}, S.~J. and {Ohm}, S. and {Olive}, J. -F. and {O{\~n}a de Wilhelmi}, E. and {Orford}, K.~J. and {Ostrowski}, M. and {Panter}, M. and {Paz Arribas}, M. and {Pedaletti}, G. and {Pelletier}, G. and {Petrucci}, P. -O. and {Pita}, S. and {P{\"u}hlhofer}, G. and {Punch}, M. and {Quirrenbach}, A. and {Raubenheimer}, B.~C. and {Raue}, M. and {Rayner}, S.~M. and {Renaud}, M. and {Rieger}, F. and {Ripken}, J. and {Rob}, L. and {Rosier-Lees}, S. and {Rowell}, G. and {Rudak}, B. and {Rulten}, C.~B. and {Ruppel}, J. and {Sahakian}, V. and {Santangelo}, A. and {Schlickeiser}, R. and {Sch{\"o}ck}, F.~M. and {Schr{\"o}der}, R. and {Schwanke}, U. and {Schwarzburg}, S. and {Schwemmer}, S. and {Shalchi}, A. and {Sikora}, M. and {Skilton}, J.~L. and {Sol}, H. and {Spangler}, D. and {Stawarz}, {\L}. and {Steenkamp}, R. and {Stegmann}, C. and {Stinzing}, F. and {Superina}, G. and {Szostek}, A. and {Tam}, P.~H. and {Tavernet}, J. -P. and {Terrier}, R. and {Tibolla}, O. and {Tluczykont}, M. and {van Eldik}, C. and {Vasileiadis}, G. and {Venter}, C. and {Venter}, L. and {Vialle}, J.~P. and {Vincent}, P. and {Vivier}, M. and {V{\"o}lk}, H.~J. and {Volpe}, F. and {Wagner}, S.~J. and {Ward}, M. and {Zdziarski}, A.~A. and {Zech}, A. and {Anderhub}, H. and {Antonelli}, L.~A. and {Antoranz}, P. and {Backes}, M. and {Baixeras}, C. and {Balestra}, S. and {Barrio}, J.~A. and {Bastieri}, D. and {Becerra Gonz{\'a}lez}, J. and {Becker}, J.~K. and {Bednarek}, W. and {Berger}, K. and {Bernardini}, E. and {Biland}, A. and {Bock}, R.~K. and {Bonnoli}, G. and {Bordas}, P. and {Tridon}, D. Borla and {Bosch-Ramon}, V. and {Bose}, D. and {Braun}, I. and {Bretz}, T. and {Britvitch}, I. and {Camara}, M. and {Carmona}, E. and {Commichau}, S. and {Contreras}, J.~L. and {Cortina}, J. and {Costado}, M.~T. and {Covino}, S. and {Curtef}, V. and {Dazzi}, F. and {De Angelis}, A. and {de Cea del Pozo}, E. and {Delgado Mendez}, C. and {De los Reyes}, R. and {De Lotto}, B. and {De Maria}, M. and {De Sabata}, F. and {Dominguez}, A. and {Dorner}, D. and {Doro}, M. and {Elsaesser}, D. and {Errando}, M. and {Ferenc}, D. and {Fern{\'a}ndez}, E. and {Firpo}, R. and {Fonseca}, M.~V. and {Font}, L. and {Galante}, N. and {Garc{\'\i}a L{\'o}pez}, R.~J. and {Garczarczyk}, M. and {Gaug}, M. and {Goebel}, F. and {Hadasch}, D. and {Hayashida}, M. and {Herrero}, A. and {Hildebrand}, D. and {H{\"o}hne-M{\"o}nch}, D. and {Hose}, J. and {Hsu}, C.~C. and {Jogler}, T. and {Kranich}, D. and {La Barbera}, A. and {Laille}, A. and {Leonardo}, E. and {Lindfors}, E. and {Lombardi}, S. and {Longo}, F. and {L{\'o}pez}, M. and {Lorenz}, E. and {Majumdar}, P. and {Maneva}, G. and {Mankuzhiyil}, N. and {Mannheim}, K. and {Maraschi}, L. and {Mariotti}, M. and {Mart{\'\i}nez}, M. and {Mazin}, D. and {Meucci}, M. and {Miranda}, J.~M. and {Mirzoyan}, R. and {Miyamoto}, H. and {Mold{\'o}n}, J. and {Moles}, M. and {Moralejo}, A. and {Nieto}, D. and {Nilsson}, K. and {Ninkovic}, J. and {Oya}, I. and {Paoletti}, R. and {Paredes}, J.~M. and {Pasanen}, M. and {Pascoli}, D. and {Pauss}, F. and {Pegna}, R.~G. and {Perez-Torres}, M.~A. and {Persic}, M. and {Peruzzo}, L. and {Prada}, F. and {Prandini}, E. and {Puchades}, N. and {Reichardt}, I. and {Rhode}, W. and {Rib{\'o}}, M. and {Rico}, J. and {Rissi}, M. and {Robert}, A. and {R{\"u}gamer}, S. and {Saggion}, A. and {Saito}, T.~Y. and {Salvati}, M. and {Sanchez-Conde}, M. and {Satalecka}, K. and {Scalzotto}, V. and {Scapin}, V. and {Schweizer}, T. and {Shayduk}, M. and {Shore}, S.~N. and {Sidro}, N. and {Sierpowska-Bartosik}, A. and {Sillanp{\"a}{\"a}}, A. and {Sitarek}, J. and {Sobczynska}, D. and {Spanier}, F. and {Stamerra}, A. and {Stark}, L.~S. and {Takalo}, L. and {Tavecchio}, F. and {Temnikov}, P. and {Tescaro}, D. and {Teshima}, M. and {Torres}, D.~F. and {Turini}, N. and {Vankov}, H. and {Wagner}, R.~M. and {Zabalza}, V. and {Zandanel}, F. and {Zanin}, R. and {Zapatero}, J. and {VERITAS Collaboration} and {VLBA 43 GHz M87 Monitoring Team} and {H.~E.~S.~S. Collaboration} and {MAGIC Collaboration}},
        title = "{Radio Imaging of the Very-High-Energy {\ensuremath{\gamma}}-Ray Emission Region in the Central Engine of a Radio Galaxy}",
      journal = {Science},
         year = 2009,
        month = jul,
       volume = {325},
       number = {5939},
        pages = {444},
          doi = {10.1126/science.1175406},
archivePrefix = {arXiv},
       eprint = {0908.0511},
 primaryClass = {astro-ph.HE},
       adsurl = {https://ui.adsabs.harvard.edu/abs/2009Sci...325..444A}
}

@ARTICLE{Perlman2005ApJ,
       author = {{Perlman}, Eric S. and {Wilson}, Andrew S.},
        title = "{The X-Ray Emissions from the M87 Jet: Diagnostics and Physical Interpretation}",
      journal = {\apj},
         year = 2005,
        month = jul,
       volume = {627},
       number = {1},
        pages = {140-155},
          doi = {10.1086/430340},
archivePrefix = {arXiv},
       eprint = {astro-ph/0503024},
 primaryClass = {astro-ph},
       adsurl = {https://ui.adsabs.harvard.edu/abs/2005ApJ...627..140P}
}

@ARTICLE{Doeleman2012Sci,
       author = {{Doeleman}, Sheperd S. and {Fish}, Vincent L. and {Schenck}, David E. and {Beaudoin}, Christopher and {Blundell}, Ray and {Bower}, Geoffrey C. and {Broderick}, Avery E. and {Chamberlin}, Richard and {Freund}, Robert and {Friberg}, Per and {Gurwell}, Mark A. and {Ho}, Paul T.~P. and {Honma}, Mareki and {Inoue}, Makoto and {Krichbaum}, Thomas P. and {Lamb}, James and {Loeb}, Abraham and {Lonsdale}, Colin and {Marrone}, Daniel P. and {Moran}, James M. and {Oyama}, Tomoaki and {Plambeck}, Richard and {Primiani}, Rurik A. and {Rogers}, Alan E.~E. and {Smythe}, Daniel L. and {SooHoo}, Jason and {Strittmatter}, Peter and {Tilanus}, Remo P.~J. and {Titus}, Michael and {Weintroub}, Jonathan and {Wright}, Melvyn and {Young}, Ken H. and {Ziurys}, Lucy M.},
        title = "{Jet-Launching Structure Resolved Near the Supermassive Black Hole in M87}",
      journal = {Science},
         year = 2012,
        month = oct,
       volume = {338},
       number = {6105},
        pages = {355},
          doi = {10.1126/science.1224768},
archivePrefix = {arXiv},
       eprint = {1210.6132},
 primaryClass = {astro-ph.HE},
       adsurl = {https://ui.adsabs.harvard.edu/abs/2012Sci...338..355D}
}

@ARTICLE{Hada2016ApJ,
       author = {{Hada}, Kazuhiro and {Kino}, Motoki and {Doi}, Akihiro and {Nagai}, Hiroshi and {Honma}, Mareki and {Akiyama}, Kazunori and {Tazaki}, Fumie and {Lico}, Rocco and {Giroletti}, Marcello and {Giovannini}, Gabriele and {Orienti}, Monica and {Hagiwara}, Yoshiaki},
        title = "{High-sensitivity 86 GHz (3.5 mm) VLBI Observations of M87: Deep Imaging of the Jet Base at a Resolution of 10 Schwarzschild Radii}",
      journal = {\apj},
         year = 2016,
        month = feb,
       volume = {817},
       number = {2},
          eid = {131},
        pages = {131},
          doi = {10.3847/0004-637X/817/2/131},
archivePrefix = {arXiv},
       eprint = {1512.03783},
 primaryClass = {astro-ph.HE},
       adsurl = {https://ui.adsabs.harvard.edu/abs/2016ApJ...817..131H}
}

@ARTICLE{Mertens2016A&A,
       author = {{Mertens}, F. and {Lobanov}, A.~P. and {Walker}, R.~C. and {Hardee}, P.~E.},
        title = "{Kinematics of the jet in M 87 on scales of 100-1000 Schwarzschild radii}",
      journal = {\aap},
         year = 2016,
        month = oct,
       volume = {595},
          eid = {A54},
        pages = {A54},
          doi = {10.1051/0004-6361/201628829},
archivePrefix = {arXiv},
       eprint = {1608.05063},
 primaryClass = {astro-ph.HE},
       adsurl = {https://ui.adsabs.harvard.edu/abs/2016A&A...595A..54M}
}

@ARTICLE{Britzen2017A&A,
       author = {{Britzen}, S. and {Fendt}, C. and {Eckart}, A. and {Karas}, V.},
        title = "{A new view on the M 87 jet origin: Turbulent loading leading to large-scale episodic wiggling}",
      journal = {\aap},
         year = 2017,
        month = may,
       volume = {601},
          eid = {A52},
        pages = {A52},
          doi = {10.1051/0004-6361/201629469},
       adsurl = {https://ui.adsabs.harvard.edu/abs/2017A&A...601A..52B}
}

@ARTICLE{Snios2019ApJ,
       author = {{Snios}, Bradford and {Nulsen}, Paul. E.~J. and {Kraft}, Ralph P. and {Cheung}, C.~C. and {Meyer}, Eileen T. and {Forman}, William R. and {Jones}, Christine and {Murray}, Stephen S.},
        title = "{Detection of Superluminal Motion in the X-Ray Jet of M87}",
      journal = {\apj},
         year = 2019,
        month = jul,
       volume = {879},
       number = {1},
          eid = {8},
        pages = {8},
          doi = {10.3847/1538-4357/ab2119},
archivePrefix = {arXiv},
       eprint = {1905.04330},
 primaryClass = {astro-ph.HE},
       adsurl = {https://ui.adsabs.harvard.edu/abs/2019ApJ...879....8S}
}

@ARTICLE{Gebhardt2011ApJ,
       author = {{Gebhardt}, Karl and {Adams}, Joshua and {Richstone}, Douglas and {Lauer}, Tod R. and {Faber}, S.~M. and {G{\"u}ltekin}, Kayhan and {Murphy}, Jeremy and {Tremaine}, Scott},
        title = "{The Black Hole Mass in M87 from Gemini/NIFS Adaptive Optics Observations}",
      journal = {\apj},
         year = 2011,
        month = mar,
       volume = {729},
       number = {2},
          eid = {119},
        pages = {119},
          doi = {10.1088/0004-637X/729/2/119},
archivePrefix = {arXiv},
       eprint = {1101.1954},
 primaryClass = {astro-ph.CO},
       adsurl = {https://ui.adsabs.harvard.edu/abs/2011ApJ...729..119G}
}

@ARTICLE{Walsh2013ApJ,
       author = {{Walsh}, Jonelle L. and {Barth}, Aaron J. and {Ho}, Luis C. and {Sarzi}, Marc},
        title = "{The M87 Black Hole Mass from Gas-dynamical Models of Space Telescope Imaging Spectrograph Observations}",
      journal = {\apj},
         year = 2013,
        month = jun,
       volume = {770},
       number = {2},
          eid = {86},
        pages = {86},
          doi = {10.1088/0004-637X/770/2/86},
archivePrefix = {arXiv},
       eprint = {1304.7273},
 primaryClass = {astro-ph.CO},
       adsurl = {https://ui.adsabs.harvard.edu/abs/2013ApJ...770...86W}
}

@ARTICLE{Macchetto1997ApJ,
       author = {{Macchetto}, F. and {Marconi}, A. and {Axon}, D.~J. and {Capetti}, A. and {Sparks}, W. and {Crane}, P.},
        title = "{The Supermassive Black Hole of M87 and the Kinematics of Its Associated Gaseous Disk}",
      journal = {\apj},
         year = 1997,
        month = nov,
       volume = {489},
       number = {2},
        pages = {579-600},
          doi = {10.1086/304823},
archivePrefix = {arXiv},
       eprint = {astro-ph/9706252},
 primaryClass = {astro-ph},
       adsurl = {https://ui.adsabs.harvard.edu/abs/1997ApJ...489..579M}
}

@ARTICLE{MAGIC2020MNRAS,
       author = {{MAGIC Collaboration} and {Acciari}, V.~A. and {Ansoldi}, S. and {Antonelli}, L.~A. and {Arbet Engels}, A. and {Arcaro}, C. and {Baack}, D. and {Babi{\'c}}, A. and {Banerjee}, B. and {Bangale}, P. and {Barres de Almeida}, U. and {Barrio}, J.~A. and {Becerra Gonz{\'a}lez}, J. and {Bednarek}, W. and {Bellizzi}, L. and {Bernardini}, E. and {Berti}, A. and {Besenrieder}, J. and {Bhattacharyya}, W. and {Bigongiari}, C. and {Biland}, A. and {Blanch}, O. and {Bonnoli}, G. and {Bo{\v{s}}njak}, {\v{Z}}. and {Busetto}, G. and {Carosi}, R. and {Ceribella}, G. and {Chai}, Y. and {Chilingaryan}, A. and {Cikota}, S. and {Colak}, S.~M. and {Colin}, U. and {Colombo}, E. and {Contreras}, J.~L. and {Cortina}, J. and {Covino}, S. and {D'Elia}, V. and {da Vela}, P. and {Dazzi}, F. and {de Angelis}, A. and {de Lotto}, B. and {Delfino}, M. and {Delgado}, J. and {Depaoli}, D. and {di Pierro}, F. and {di Venere}, L. and {Do Souto Espi{\~n}eira}, E. and {Dominis Prester}, D. and {Donini}, A. and {Dorner}, D. and {Doro}, M. and {Elsaesser}, D. and {Fallah Ramazani}, V. and {Fattorini}, A. and {Fern{\'a}ndez-Barral}, A. and {Ferrara}, G. and {Fidalgo}, D. and {Foffano}, L. and {Fonseca}, M.~V. and {Font}, L. and {Fruck}, C. and {Fukami}, S. and {Garc{\'\i}a L{\'o}pez}, R.~J. and {Garczarczyk}, M. and {Gasparyan}, S. and {Gaug}, M. and {Giglietto}, N. and {Giordano}, F. and {Godinovi{\'c}}, N. and {Green}, D. and {Guberman}, D. and {Hadasch}, D. and {Hahn}, A. and {Herrera}, J. and {Hoang}, J. and {Hrupec}, D. and {H{\"u}tten}, M. and {Inada}, T. and {Inoue}, S. and {Ishio}, K. and {Iwamura}, Y. and {Jouvin}, L. and {Kerszberg}, D. and {Kubo}, H. and {Kushida}, J. and {Lamastra}, A. and {Lelas}, D. and {Leone}, F. and {Lindfors}, E. and {Lombardi}, S. and {Longo}, F. and {L{\'o}pez}, M. and {L{\'o}pez-Coto}, R. and {L{\'o}pez-Oramas}, A. and {Loporchio}, S. and {Machado de Oliveira Fraga}, B. and {Maggio}, C. and {Majumdar}, P. and {Makariev}, M. and {Mallamaci}, M. and {Maneva}, G. and {Manganaro}, M. and {Mannheim}, K. and {Maraschi}, L. and {Mariotti}, M. and {Mart{\'\i}nez}, M. and {Masuda}, S. and {Mazin}, D. and {Mi{\'c}anovi{\'c}}, S. and {Miceli}, D. and {Minev}, M. and {Miranda}, J.~M. and {Mirzoyan}, R. and {Molina}, E. and {Moralejo}, A. and {Morcuende}, D. and {Moreno}, V. and {Moretti}, E. and {Munar-Adrover}, P. and {Neustroev}, V. and {Nigro}, C. and {Nilsson}, K. and {Ninci}, D. and {Nishijima}, K. and {Noda}, K. and {Nogu{\'e}s}, L. and {N{\"o}the}, M. and {Nozaki}, S. and {Paiano}, S. and {Palacio}, J. and {Palatiello}, M. and {Paneque}, D. and {Paoletti}, R. and {Paredes}, J.~M. and {Pe{\~n}il}, P. and {Peresano}, M. and {Persic}, M. and {Prada Moroni}, P.~G. and {Prandini}, E. and {Puljak}, I. and {Rhode}, W. and {Rib{\'o}}, M. and {Rico}, J. and {Righi}, C. and {Rugliancich}, A. and {Saha}, L. and {Sahakyan}, N. and {Saito}, T. and {Sakurai}, S. and {Satalecka}, K. and {Schmidt}, K. and {Schweizer}, T. and {Sitarek}, J. and {{\v{S}}nidari{\'c}}, I. and {Sobczynska}, D. and {Somero}, A. and {Stamerra}, A. and {Strom}, D. and {Strzys}, M. and {Suda}, Y. and {Suri{\'c}}, T. and {Takahashi}, M. and {Tavecchio}, F. and {Temnikov}, P. and {Terzi{\'c}}, T. and {Teshima}, M. and {Torres-Alb{\`a}}, N. and {Tosti}, L. and {Tsujimoto}, S. and {Vagelli}, V. and {van Scherpenberg}, J. and {Vanzo}, G. and {Acosta}, M. Vazquez and {Vigorito}, C.~F. and {Vitale}, V. and {Vovk}, I. and {Will}, M. and {Zari{\'c}}, D. and {Asano}, K. and {Hada}, K. and {Harris}, D.~E. and {Giroletti}, M. and {Jermak}, H.~E. and {Madrid}, J.~P. and {Massaro}, F. and {Richter}, S. and {Spanier}, F. and {Steele}, I.~A. and {Walker}, R.~C.},
        title = "{Monitoring of the radio galaxy M 87 during a low-emission state from 2012 to 2015 with MAGIC}",
      journal = {\mnras},
         year = 2020,
        month = mar,
       volume = {492},
       number = {4},
        pages = {5354-5365},
          doi = {10.1093/mnras/staa014},
archivePrefix = {arXiv},
       eprint = {2001.01643},
 primaryClass = {astro-ph.HE},
       adsurl = {https://ui.adsabs.harvard.edu/abs/2020MNRAS.492.5354M}
}

@ARTICLE{Sahayanathan2008MNRAS,
       author = {{Sahayanathan}, S.},
        title = "{A two-zone synchrotron model for the knots in the M87 jet}",
      journal = {\mnras},
         year = 2008,
        month = jul,
       volume = {388},
       number = {1},
        pages = {L49-L53},
          doi = {10.1111/j.1745-3933.2008.00497.x},
archivePrefix = {arXiv},
       eprint = {0805.2842},
 primaryClass = {astro-ph},
       adsurl = {https://ui.adsabs.harvard.edu/abs/2008MNRAS.388L..49S}
}

@INPROCEEDINGS{Bednarek1999ptep,
       author = {{Bednarek}, W. {\l} Odek},
        title = "{Gamma-Rays from Collisions of Compact Objects with AGN Jets?}",
    booktitle = {Plasma Turbulence and Energetic Particles in Astrophysics},
         year = 1999,
       editor = {{Ostrowski}, Micha{\l} and {Schlickeiser}, Reinhard},
        month = dec,
        pages = {360-365},
          doi = {10.48550/arXiv.astro-ph/9911271},
archivePrefix = {arXiv},
       eprint = {astro-ph/9911271},
 primaryClass = {astro-ph},
       adsurl = {https://ui.adsabs.harvard.edu/abs/1999ptep.proc..360B}
}

@ARTICLE{Davis2001ApJ,
       author = {{Davis}, John E.},
        title = "{Event Pileup in Charge-coupled Devices}",
      journal = {\apj},
         year = 2001,
        month = nov,
       volume = {562},
       number = {1},
        pages = {575-582},
          doi = {10.1086/323488},
       adsurl = {https://ui.adsabs.harvard.edu/abs/2001ApJ...562..575D}
}

@ARTICLE{Vieyro2017A&A,
       author = {{Vieyro}, Florencia L. and {Torres-Alb{\`a}}, N{\'u}ria and {Bosch-Ramon}, Valent{\'\i}},
        title = "{Collective non-thermal emission from an extragalactic jet interacting with stars}",
      journal = {\aap},
         year = 2017,
        month = aug,
       volume = {604},
          eid = {A57},
        pages = {A57},
          doi = {10.1051/0004-6361/201630333},
archivePrefix = {arXiv},
       eprint = {1704.01919},
 primaryClass = {astro-ph.HE},
       adsurl = {https://ui.adsabs.harvard.edu/abs/2017A&A...604A..57V}
}

@ARTICLE{Fan2008ApJ,
       author = {{Fan}, Zhong-Hui and {Liu}, Siming and {Wang}, Jian-Min and {Fryer}, Christopher L. and {Li}, Hui},
        title = "{Stochastic Acceleration in the Western Hot Spot of Pictor A}",
      journal = {\apjl},
         year = 2008,
        month = feb,
       volume = {673},
       number = {2},
        pages = {L139},
          doi = {10.1086/528372},
archivePrefix = {arXiv},
       eprint = {0711.4794},
 primaryClass = {astro-ph},
       adsurl = {https://ui.adsabs.harvard.edu/abs/2008ApJ...673L.139F}
}

@ARTICLE{Dayal2018PhR,
       author = {{Dayal}, Pratika and {Ferrara}, Andrea},
        title = "{Early galaxy formation and its large-scale effects}",
      journal = {\physrep},
         year = 2018,
        month = dec,
       volume = {780},
        pages = {1-64},
          doi = {10.1016/j.physrep.2018.10.002},
archivePrefix = {arXiv},
       eprint = {1809.09136},
 primaryClass = {astro-ph.GA},
       adsurl = {https://ui.adsabs.harvard.edu/abs/2018PhR...780....1D}
}

@ARTICLE{Cao2024ApJ,
       author = {{Cao}, Zhen and {Aharonian}, F. and {Axikegu} and {Bai}, Y.~X. and {Bao}, Y.~W. and {Bastieri}, D. and {Bi}, X.~J. and {Bi}, Y.~J. and {Bian}, W. and {Bukevich}, A.~V. and {Cao}, Q. and {Cao}, W.~Y. and {Cao}, Zhe and {Chang}, J. and {Chang}, J.~F. and {Chen}, A.~M. and {Chen}, E.~S. and {Chen}, H.~X. and {Chen}, Liang and {Chen}, Lin and {Chen}, Long and {Chen}, M.~J. and {Chen}, M.~L. and {Chen}, Q.~H. and {Chen}, S. and {Chen}, S.~H. and {Chen}, S.~Z. and {Chen}, T.~L. and {Chen}, Y. and {Cheng}, N. and {Cheng}, Y.~D. and {Chu}, M.~C. and {Cui}, M.~Y. and {Cui}, S.~W. and {Cui}, X.~H. and {Cui}, Y.~D. and {Dai}, B.~Z. and {Dai}, H.~L. and {Dai}, Z.~G. and {Danzengluobu} and {Dong}, X.~Q. and {Duan}, K.~K. and {Fan}, J.~H. and {Fan}, Y.~Z. and {Fang}, J. and {Fang}, J.~H. and {Fang}, K. and {Feng}, C.~F. and {Feng}, H. and {Feng}, L. and {Feng}, S.~H. and {Feng}, X.~T. and {Feng}, Y. and {Feng}, Y.~L. and {Gabici}, S. and {Gao}, B. and {Gao}, C.~D. and {Gao}, Q. and {Gao}, W. and {Gao}, W.~K. and {Ge}, M.~M. and {Ge}, T.~T. and {Geng}, L.~S. and {Giacinti}, G. and {Gong}, G.~H. and {Gou}, Q.~B. and {Gu}, M.~H. and {Guo}, F.~L. and {Guo}, J. and {Guo}, X.~L. and {Guo}, Y.~Q. and {Guo}, Y.~Y. and {Han}, Y.~A. and {Hannuksela}, O.~A. and {Hasan}, M. and {He}, H.~H. and {He}, H.~N. and {He}, J.~Y. and {He}, Y. and {Hor}, Y.~K. and {Hou}, B.~W. and {Hou}, C. and {Hou}, X. and {Hu}, H.~B. and {Hu}, Q. and {Hu}, S.~C. and {Huang}, C. and {Huang}, D.~H. and {Huang}, T.~Q. and {Huang}, W.~J. and {Huang}, X.~T. and {Huang}, X.~Y. and {Huang}, Y. and {Huang}, Y.~Y. and {Ji}, X.~L. and {Jia}, H.~Y. and {Jia}, K. and {Jiang}, H.~B. and {Jiang}, K. and {Jiang}, X.~W. and {Jiang}, Z.~J. and {Jin}, M. and {Kang}, M.~M. and {Karpikov}, I. and {Khangulyan}, D. and {Kuleshov}, D. and {Kurinov}, K. and {Li}, B.~B. and {Li}, C.~M. and {Li}, Cheng and {Li}, Cong and {Li}, D. and {Li}, F. and {Li}, H.~B. and {Li}, H.~C. and {Li}, Jian and {Li}, Jie and {Li}, K. and {Li}, S.~D. and {Li}, W.~L. and {Li}, W.~L. and {Li}, X.~R. and {Li}, Xin and {Li}, Y.~Z. and {Li}, Zhe and {Li}, Zhuo and {Liang}, E.~W. and {Liang}, Y.~F. and {Lin}, S.~J. and {Liu}, B. and {Liu}, C. and {Liu}, D. and {Liu}, D.~B. and {Liu}, H. and {Liu}, H.~D. and {Liu}, J. and {Liu}, J.~L. and {Liu}, M.~Y. and {Liu}, R.~Y. and {Liu}, S.~M. and {Liu}, W. and {Liu}, Y. and {Liu}, Y.~N. and {Luo}, Q. and {Luo}, Y. and {Lv}, H.~K. and {Ma}, B.~Q. and {Ma}, L.~L. and {Ma}, X.~H. and {Mao}, J.~R. and {Min}, Z. and {Mitthumsiri}, W. and {Mu}, H.~J. and {Nan}, Y.~C. and {Neronov}, A. and {Ng}, K.~C.~Y. and {Ou}, L.~J. and {Pattarakijwanich}, P. and {Pei}, Z.~Y. and {Qi}, J.~C. and {Qi}, M.~Y. and {Qiao}, B.~Q. and {Qin}, J.~J. and {Raza}, A. and {Ruffolo}, D. and {S{\'a}iz}, A. and {Saeed}, M. and {Semikoz}, D. and {Shao}, L. and {Shchegolev}, O. and {Sheng}, X.~D. and {Shu}, F.~W. and {Song}, H.~C. and {Stenkin}, Yu. V. and {Stepanov}, V. and {Su}, Y. and {Sun}, D.~X. and {Sun}, Q.~N. and {Sun}, X.~N. and {Sun}, Z.~B. and {Takata}, J. and {Tam}, P.~H.~T. and {Tang}, Q.~W. and {Tang}, R. and {Tang}, Z.~B. and {Tian}, W.~W. and {Wan}, L.~H. and {Wang}, C. and {Wang}, C.~B. and {Wang}, G.~W. and {Wang}, H.~G. and {Wang}, H.~H. and {Wang}, J.~C. and {Wang}, Kai and {Wang}, Kai and {Wang}, L.~P. and {Wang}, L.~Y. and {Wang}, P.~H. and {Wang}, R. and {Wang}, W.},
        title = "{Detection of Very High-energy Gamma-Ray Emission from the Radio Galaxy M87 with LHAASO}",
      journal = {\apjl},
         year = 2024,
        month = nov,
       volume = {975},
       number = {2},
          eid = {L44},
        pages = {L44},
          doi = {10.3847/2041-8213/ad8921},
archivePrefix = {arXiv},
       eprint = {2410.15353},
 primaryClass = {astro-ph.HE},
       adsurl = {https://ui.adsabs.harvard.edu/abs/2024ApJ...975L..44C}
}

@ARTICLE{Reynolds1996MNRAS,
       author = {{Reynolds}, C.~S. and {Di Matteo}, T. and {Fabian}, A.~C. and {Hwang}, U. and {Canizares}, C.~R.},
        title = "{The `quiescent' black hole in M87}",
      journal = {\mnras},
         year = 1996,
        month = dec,
       volume = {283},
       number = {4},
        pages = {L111-L116},
          doi = {10.1093/mnras/283.4.L111},
archivePrefix = {arXiv},
       eprint = {astro-ph/9610097},
 primaryClass = {astro-ph},
       adsurl = {https://ui.adsabs.harvard.edu/abs/1996MNRAS.283L.111R}
}

@ARTICLE{Gasperin2012A&A,
       author = {{de Gasperin}, F. and {Orr{\'u}}, E. and {Murgia}, M. and {Merloni}, A. and {Falcke}, H. and {Beck}, R. and {Beswick}, R. and {B{\^\i}rzan}, L. and {Bonafede}, A. and {Br{\"u}ggen}, M. and {Brunetti}, G. and {Chy{\.z}y}, K. and {Conway}, J. and {Croston}, J.~H. and {En{\ss}lin}, T. and {Ferrari}, C. and {Heald}, G. and {Heidenreich}, S. and {Jackson}, N. and {Macario}, G. and {McKean}, J. and {Miley}, G. and {Morganti}, R. and {Offringa}, A. and {Pizzo}, R. and {Rafferty}, D. and {R{\"o}ttgering}, H. and {Shulevski}, A. and {Steinmetz}, M. and {Tasse}, C. and {van der Tol}, S. and {van Driel}, W. and {van Weeren}, R.~J. and {van Zwieten}, J.~E. and {Alexov}, A. and {Anderson}, J. and {Asgekar}, A. and {Avruch}, M. and {Bell}, M. and {Bell}, M.~R. and {Bentum}, M. and {Bernardi}, G. and {Best}, P. and {Breitling}, F. and {Broderick}, J.~W. and {Butcher}, A. and {Ciardi}, B. and {Dettmar}, R.~J. and {Eisloeffel}, J. and {Frieswijk}, W. and {Gankema}, H. and {Garrett}, M. and {Gerbers}, M. and {Griessmeier}, J.~M. and {Gunst}, A.~W. and {Hassall}, T.~E. and {Hessels}, J. and {Hoeft}, M. and {Horneffer}, A. and {Karastergiou}, A. and {K{\"o}hler}, J. and {Koopman}, Y. and {Kuniyoshi}, M. and {Kuper}, G. and {Maat}, P. and {Mann}, G. and {Mevius}, M. and {Mulcahy}, D.~D. and {Munk}, H. and {Nijboer}, R. and {Noordam}, J. and {Paas}, H. and {Pandey}, M. and {Pandey}, V.~N. and {Polatidis}, A. and {Reich}, W. and {Schoenmakers}, A.~P. and {Sluman}, J. and {Smirnov}, O. and {Sobey}, C. and {Stappers}, B. and {Swinbank}, J. and {Tagger}, M. and {Tang}, Y. and {van Bemmel}, I. and {van Cappellen}, W. and {van Duin}, A.~P. and {van Haarlem}, M. and {van Leeuwen}, J. and {Vermeulen}, R. and {Vocks}, C. and {White}, S. and {Wise}, M. and {Wucknitz}, O. and {Zarka}, P.},
        title = "{M 87 at metre wavelengths: the LOFAR picture}",
      journal = {\aap},
         year = 2012,
        month = nov,
       volume = {547},
          eid = {A56},
        pages = {A56},
          doi = {10.1051/0004-6361/201220209},
archivePrefix = {arXiv},
       eprint = {1210.1346},
 primaryClass = {astro-ph.GA},
       adsurl = {https://ui.adsabs.harvard.edu/abs/2012A&A...547A..56D}
}

@ARTICLE{Mo2016A&A,
       author = {{Mo{\'s}cibrodzka}, Monika and {Falcke}, Heino and {Shiokawa}, Hotaka},
        title = "{General relativistic magnetohydrodynamical simulations of the jet in M 87}",
      journal = {\aap},
         year = 2016,
        month = feb,
       volume = {586},
          eid = {A38},
        pages = {A38},
          doi = {10.1051/0004-6361/201526630},
archivePrefix = {arXiv},
       eprint = {1510.07243},
 primaryClass = {astro-ph.HE},
       adsurl = {https://ui.adsabs.harvard.edu/abs/2016A&A...586A..38M}
}

@ARTICLE{Levinson2017MNRAS,
       author = {{Levinson}, A. and {Globus}, N.},
        title = "{Reconfinement of highly magnetized jets: implications for HST-1 in M87}",
      journal = {\mnras},
         year = 2017,
        month = feb,
       volume = {465},
       number = {2},
        pages = {1608-1612},
          doi = {10.1093/mnras/stw2902},
archivePrefix = {arXiv},
       eprint = {1609.01091},
 primaryClass = {astro-ph.HE},
       adsurl = {https://ui.adsabs.harvard.edu/abs/2017MNRAS.465.1608L}
}

@ARTICLE{Fedorenko1996A&A,
       author = {{Fedorenko}, V.~N. and {Courvoisier}, T.~J. -L.},
        title = "{A model for radio/optical jets.}",
      journal = {\aap},
         year = 1996,
        month = mar,
       volume = {307},
        pages = {347-358},
       adsurl = {https://ui.adsabs.harvard.edu/abs/1996A&A...307..347F}
}

@ARTICLE{Torres2019A&A,
       author = {{Torres-Alb{\`a}}, N{\'u}ria and {Bosch-Ramon}, Valent{\'\i}},
        title = "{Gamma rays from red giant wind bubbles entering the jets of elliptical host blazars}",
      journal = {\aap},
         year = 2019,
        month = mar,
       volume = {623},
          eid = {A91},
        pages = {A91},
          doi = {10.1051/0004-6361/201833697},
archivePrefix = {arXiv},
       eprint = {1902.05008},
 primaryClass = {astro-ph.HE},
       adsurl = {https://ui.adsabs.harvard.edu/abs/2019A&A...623A..91T}
}

@ARTICLE{Richardson1972JOSA,
       author = {{Richardson}, William Hadley},
        title = "{Bayesian-Based Iterative Method of Image Restoration}",
      journal = {Journal of the Optical Society of America (1917-1983)},
         year = 1972,
        month = jan,
       volume = {62},
       number = {1},
        pages = {55},
       adsurl = {https://ui.adsabs.harvard.edu/abs/1972JOSA...62...55R}
}

@ARTICLE{Lucy1974AJ,
       author = {{Lucy}, L.~B.},
        title = "{An iterative technique for the rectification of observed distributions}",
      journal = {\aj},
         year = 1974,
        month = jun,
       volume = {79},
        pages = {745},
          doi = {10.1086/111605},
       adsurl = {https://ui.adsabs.harvard.edu/abs/1974AJ.....79..745L}
}

@ARTICLE{Fichet2025A&A,
       author = {{Fichet de Clairfontaine}, G. and {Perucho}, M. and {Mart{\'\i}}, J.~M. and {Kovalev}, Y.~Y.},
        title = "{Dynamic and radiative implications of jet{\textendash}star interactions in AGN jets}",
      journal = {\aap},
         year = 2025,
        month = jan,
       volume = {693},
          eid = {A270},
        pages = {A270},
          doi = {10.1051/0004-6361/202451914},
archivePrefix = {arXiv},
       eprint = {2412.07945},
 primaryClass = {astro-ph.HE},
       adsurl = {https://ui.adsabs.harvard.edu/abs/2025A&A...693A.270F}
}

@ARTICLE{Longo2025A&A,
       author = {{Longo}, B. and {Perucho}, M. and {Bosch-Ramon}, V. and {Mart{\'\i}}, J.~M. and {Fichet de Clairfontaine}, G.},
        title = "{Relativistic hydrodynamics simulations of supernova explosions within extragalactic jets}",
      journal = {\aap},
         year = 2025,
        month = dec,
       volume = {704},
          eid = {A172},
        pages = {A172},
          doi = {10.1051/0004-6361/202555849},
archivePrefix = {arXiv},
       eprint = {2510.04570},
 primaryClass = {astro-ph.HE},
       adsurl = {https://ui.adsabs.harvard.edu/abs/2025A&A...704A.172L}
}

@ARTICLE{Perucho2019Galax,
       author = {{Perucho}, Manel},
        title = "{Dissipative Processes and Their Role in the Evolution of Radio Galaxies}",
      journal = {Galaxies},
         year = 2019,
        month = jul,
       volume = {7},
       number = {3},
          eid = {70},
        pages = {70},
          doi = {10.3390/galaxies7030070},
archivePrefix = {arXiv},
       eprint = {1907.13599},
 primaryClass = {astro-ph.HE},
       adsurl = {https://ui.adsabs.harvard.edu/abs/2019Galax...7...70P}
}

@ARTICLE{Marshall2002ApJ,
       author = {{Marshall}, H.~L. and {Miller}, B.~P. and {Davis}, D.~S. and {Perlman}, E.~S. and {Wise}, M. and {Canizares}, C.~R. and {Harris}, D.~E.},
        title = "{A High-Resolution X-Ray Image of the Jet in M87}",
      journal = {\apj},
         year = 2002,
        month = jan,
       volume = {564},
       number = {2},
        pages = {683-687},
          doi = {10.1086/324396},
archivePrefix = {arXiv},
       eprint = {astro-ph/0109160},
 primaryClass = {astro-ph},
       adsurl = {https://ui.adsabs.harvard.edu/abs/2002ApJ...564..683M}
}

@ARTICLE{Begelman1984RvMP,
       author = {{Begelman}, Mitchell C. and {Blandford}, Roger D. and {Rees}, Martin J.},
        title = "{Theory of extragalactic radio sources}",
      journal = {Reviews of Modern Physics},
         year = 1984,
        month = apr,
       volume = {56},
       number = {2},
        pages = {255-351},
          doi = {10.1103/RevModPhys.56.255},
       adsurl = {https://ui.adsabs.harvard.edu/abs/1984RvMP...56..255B}
}

@ARTICLE{Dermer1995ApJ,
       author = {{Dermer}, Charles D.},
        title = "{On the Beaming Statistics of Gamma-Ray Sources}",
      journal = {\apjl},
         year = 1995,
        month = jun,
       volume = {446},
        pages = {L63},
          doi = {10.1086/187931},
       adsurl = {https://ui.adsabs.harvard.edu/abs/1995ApJ...446L..63D}
}

@ARTICLE{Jord2005ApJ,
       author = {{Jord{\'a}n}, Andr{\'e}s and {C{\^o}t{\'e}}, Patrick and {Blakeslee}, John P. and {Ferrarese}, Laura and {McLaughlin}, Dean E. and {Mei}, Simona and {Peng}, Eric W. and {Tonry}, John L. and {Merritt}, David and {Milosavljevi{\'c}}, Milo{\v{s}} and {Sarazin}, Craig L. and {Sivakoff}, Gregory R. and {West}, Michael J.},
        title = "{The ACS Virgo Cluster Survey. X. Half-Light Radii of Globular Clusters in Early-Type Galaxies: Environmental Dependencies and a Standard Ruler for Distance Estimation}",
      journal = {\apj},
         year = 2005,
        month = dec,
       volume = {634},
       number = {2},
        pages = {1002-1019},
          doi = {10.1086/497092},
archivePrefix = {arXiv},
       eprint = {astro-ph/0508219},
 primaryClass = {astro-ph},
       adsurl = {https://ui.adsabs.harvard.edu/abs/2005ApJ...634.1002J}
}

@ARTICLE{Blakeslee2009ApJ,
       author = {{Blakeslee}, John P. and {Jord{\'a}n}, Andr{\'e}s and {Mei}, Simona and {C{\^o}t{\'e}}, Patrick and {Ferrarese}, Laura and {Infante}, Leopoldo and {Peng}, Eric W. and {Tonry}, John L. and {West}, Michael J.},
        title = "{The ACS Fornax Cluster Survey. V. Measurement and Recalibration of Surface Brightness Fluctuations and a Precise Value of the Fornax-Virgo Relative Distance}",
      journal = {\apj},
         year = 2009,
        month = mar,
       volume = {694},
       number = {1},
        pages = {556-572},
          doi = {10.1088/0004-637X/694/1/556},
archivePrefix = {arXiv},
       eprint = {0901.1138},
 primaryClass = {astro-ph.CO},
       adsurl = {https://ui.adsabs.harvard.edu/abs/2009ApJ...694..556B}
}

@ARTICLE{Caprioli2014ApJ,
       author = {{Caprioli}, D. and {Spitkovsky}, A.},
        title = "{Simulations of Ion Acceleration at Non-relativistic Shocks. I. Acceleration Efficiency}",
      journal = {\apj},
         year = 2014,
        month = mar,
       volume = {783},
       number = {2},
          eid = {91},
        pages = {91},
          doi = {10.1088/0004-637X/783/2/91},
archivePrefix = {arXiv},
       eprint = {1310.2943},
 primaryClass = {astro-ph.HE},
       adsurl = {https://ui.adsabs.harvard.edu/abs/2014ApJ...783...91C}
}

@ARTICLE{Dermer2002ApJ,
       author = {{Dermer}, Charles D. and {Schlickeiser}, Reinhard},
        title = "{Transformation Properties of External Radiation Fields, Energy-Loss Rates and Scattered Spectra, and a Model for Blazar Variability}",
      journal = {\apj},
         year = 2002,
        month = aug,
       volume = {575},
       number = {2},
        pages = {667-686},
          doi = {10.1086/341431},
archivePrefix = {arXiv},
       eprint = {astro-ph/0202280},
 primaryClass = {astro-ph},
       adsurl = {https://ui.adsabs.harvard.edu/abs/2002ApJ...575..667D}
}

@ARTICLE{Drury1983RPPh,
       author = {{Drury}, L. Oc.},
        title = "{REVIEW ARTICLE: An introduction to the theory of diffusive shock acceleration of energetic particles in tenuous plasmas}",
      journal = {Reports on Progress in Physics},
         year = 1983,
        month = aug,
       volume = {46},
       number = {8},
        pages = {973-1027},
          doi = {10.1088/0034-4885/46/8/002},
       adsurl = {https://ui.adsabs.harvard.edu/abs/1983RPPh...46..973D}
}

@BOOK{Rybicki1979rpa..book.....R,
       author = {{Rybicki}, George B. and {Lightman}, Alan P.},
        title = "{Radiative processes in astrophysics}",
         year = 1979,
       adsurl = {https://ui.adsabs.harvard.edu/abs/1979rpa..book.....R}
}

@BOOK{Aharonian2004vhec.book.....A,
       author = {{Aharonian}, Felix A.},
        title = "{Very high energy cosmic gamma radiation : a crucial window on the extreme Universe}",
         year = 2004,
          doi = {10.1142/4657},
       adsurl = {https://ui.adsabs.harvard.edu/abs/2004vhec.book.....A}
}

\appendix 

\section{Data analysis of \chandra} \label{appendix_data}
\ad{The \chandra\ \xray Observatory launched in 1999, provides unprecedented angular resolution ($<0.5\rm \arcsec$) \xray imaging and spectroscopy in the energy range $0.1-10$ keV \citep{Weisskopf2002}. 
The Science Instrument Module of \chandra \ has two focal plane instruments, the Advanced CCD Imaging Spectrometer (ACIS) and the High Resolution Camera (HRC). 
The ACIS module is used for spectral analysis.
In this paper, the spectral extraction is performed using the CIAO (v4.16) software and the \chandra \ Calibration Database (CALDB, v4.11.2). 
The spectral analysis is performed using {\it Sherpa}\footnote{\url{https://cxc.harvard.edu/sherpa/threads/index.html}} tool. }
%


\ad{The \xray data of M 87 are all from the ACIS module of \chandra\ \xray Observatory. 
Owing to the cumulative exposure and the enhanced software tools of \chandra, we perform an improved analysis for M 87 to derive more accurate spectrometric information.
The observational data for M 87 span from July 29, 2000 (ObsID 352) to May 6, 2022 (ObsID 25369), with a total exposure time exceeding 1,665 kiloseconds.
We analyze the \chandra-ACIS data following the guidance of $Science\ Threads$\footnote{\url{https://cxc.harvard.edu/ciao/threads/index.html}}. 
In order to reduce the deviations caused by the position offsets of different observations, we perform astrometric corrections. 
The counts image, the exposure map, and the weighted point spread function (PSF) map are produced by performing {\it fluximage} and {\it mkpsfmap} tools, respectively. 
We obtain the locations of target sources using the {\it wavdetect} tool. 
We select the longest exposure observation (ObsID 18838) as a reference, and perform the cross-matching between the 
reference observation and the others. 
We use {\it wcs\_match} to produce a transform 
matrix and {\it wcs\_update} tool to update the coordinates of the shorter observation. }



\begin{figure}
    \raggedright 
    \begin{minipage}{0.48\textwidth} 
        \raggedright
        \includegraphics[width=\textwidth]{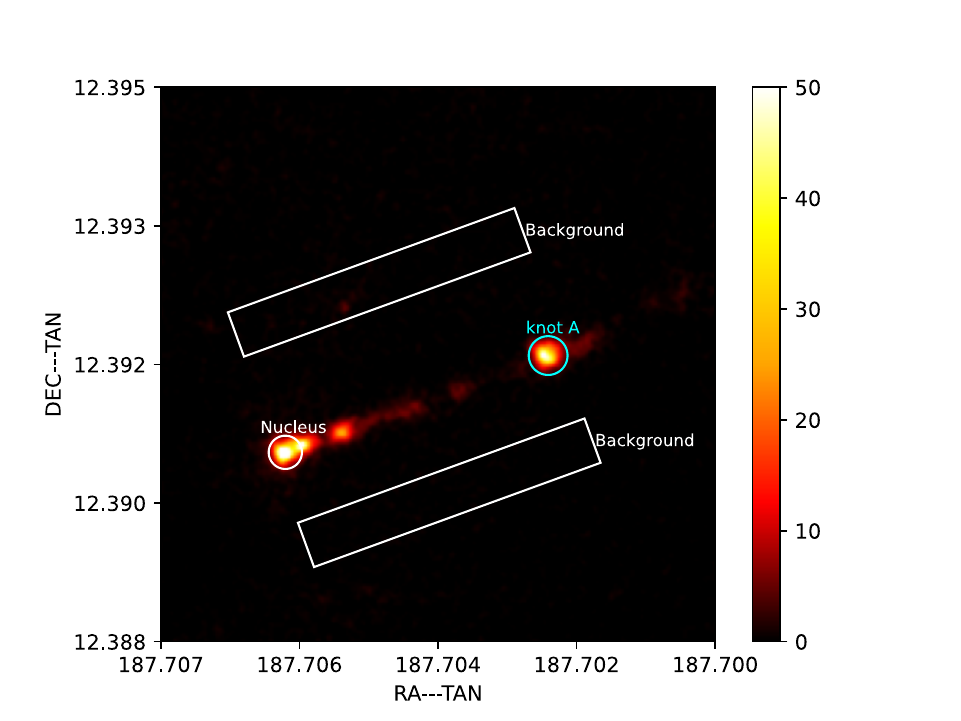} 
        \caption{Deconvolved ACIS image of the \chandra\ observation on July 30, 2000 (ObsID 1808), binned in 0.1\arcsec per pixel. The cyan circle indicates the knot A region. The units of RA-- TAN and DEC--TAN are degrees.}
        \label{figure:jet}
    \end{minipage}
\end{figure}
To avoid pile-up \citep{Davis2001ApJ}, we select observations with a frame time of 0.4/0.5 s to perform the spectral analysis. 
The observations with the frame time 0.4/0.5 s are listed in Table \ref{table:chandra_obs}. 
We perform PSF deconvolution on the \xray data of M 87 to determine an optimal energy. 
For this purpose, we divide the 0.5$-$7.0 keV band into three uniformly spaced logarithmic bins and conduct independent Richardson-Lucy deconvolutions using the \textit{arestore} tool \citep{Richardson1972JOSA,Lucy1974AJ} with PSF models generated by the $simulate\_psf$ tool for each bin. 
Since the deconvolved images show no significant morphological variations across these energy bins, we select 2.3 keV (the logarithmic mid-point of this energy band) as the characteristic energy to obtain the final deconvolved \xray images. 
The resulting image is shown in Figure \ref{figure:jet}, which presents the \chandra\ observation on July 30, 2000 (ObsID 1808) in $0.1 - 10$ keV, binned in $0.1\arcsec$ per pixel. 
We define the source region of knot A  based on the \xray positions in Table 1 of \citep{Perlman2005ApJ}. 
The intrinsic physical scales of knot A are defined by its 90\% encircled energy radius. 
The corresponding radius of knot A (R.A. = $12^{\rm h} 30^{\rm m} 48^{\rm s}.621$, Decl. = $12^{\rm h} 23^{\rm m} 32^{\rm s}.29$, J2000, at a distance from the nucleus of 12.6$\arcsec$) is defined as $R_{\rm rad, A} \approx 0.77 \arcsec \times 78\ \rm pc/\arcsec \approx 60$ pc. 
We assume a jet half-opening angle of 0.1 rad for M 87 jet, the jet width at knot A is derived to be $R_{\rm j, A}\sim 100$ pc \citep{Bosch-Ramon2015A&A}. 
We perform aperture photometry using {\it specextract} on knot A for the spectral analysis. 
The background regions are defined as the two white rectangles in Figure \ref{figure:jet}. 
We use the {\it sherpa} package to perform the broadband fitting of multi-observations simultaneously with a single power-law plus the Galactic absorption model.
The flux of knot A is extracted in the 0.5$-$7.0 keV energy band, and it is divided into three energy bins. %
We do not find evidence for significant deviation of the flux and indices for knot A if \nh\, is kept frozen, thus, we keep \nh\, free. 
The \xray flux and reduced chi-square $\chi^{2}$ are listed in Table~\ref{tab:Xray_data}. 
The errors of flux and photon index are calculated at a 90\% confidence level. 
For knot A, we obtain the spectral index of $1.4^{+0.02}_{-0.03}$, with an energy flux of $(1.97-3.77)\times10^{-13}\ \rm erg\ \rm cm^{-2}\ s^{-1}$ and  $\nh$ of $\leq 2.12\times10^{19}\ \rm cm^{-2}$. 
We calculate the \xray luminosities ($L_{\rm X}$) of knot A based on the average flux ($\overline{\nu F_{\rm \nu}}$) in the 0.5$-$7.0 keV energy band, i.e. $L_{\rm X}=4\pi D^{2}_{\rm L} \overline{\nu F_{\rm \nu}}$. 
The luminosities in the 0.5$-$7.0 keV energy band are calculated to be $\sim 10^{40}\ \rm erg\ s^{-1}$ for knot A. 
We compute the radio-to-optical spectral index $\alpha_{\rm RO}$ using archival data from \citet{Perlman2001}.
The discrepancy between the radio-to-optical spectral index ($\alpha_{\rm RO}$) and the X-ray spectral index ($\alpha_{\rm X}$), as shown in Table~\ref{tab:Xray_data}, implies that a single electron population cannot account for the broadband radio-to-X-ray SED.

\begin{table*} 
    \raggedright 

    \begin{threeparttable}
        \raggedright
        \caption{The results for the X-ray spectral fitting of knot A for different energy bands.}
        \label{tab:Xray_data} 
        
        \renewcommand{\arraystretch}{1.6}
        \begin{tabular}{c|ccc|ccc|c}
        \hline\hline
        \multirow{2}{*}{} & $\nu F_{\nu}$(0.5$-$1.2 keV)& $\nu F_{\nu}$(1.2$-$2.9 keV) & $\nu F_{\nu}$(2.9$-$7.0 keV) &  $\nh$ \tnote{a}  & $\alpha_{\rm X} $ & reduced $\chi^{2}$&$\alpha_{\rm RO}$ \\
         &\multicolumn{3}{c|}{[$\times10^{-14} \ \rm erg\ \rm cm^{-2}\ s^{-1}$]} & [$\times 10^{19}\ \rm cm^{-2}$]& & & \\
        \hline
        M 87 - knot A & 37.71$\pm$0.08 & 27.43$\pm$0.07 & 19.69$\pm$0.11 &  $<2.12$ &$1.44^{+0.02}_{-0.03}$ & 0.86 & 0.67\\ 
        \hline\hline
        \end{tabular}
        \begin{tablenotes}
            \footnotesize
            \item[a] Hydrogen-absorbing column density. \\
            Spectral index $\alpha$ and flux are expressed as $F_\nu \propto \nu^{-\alpha}$.
        \end{tablenotes}
    \end{threeparttable}

    \vspace{1.5cm} 

    \includegraphics[height=13cm, keepaspectratio]{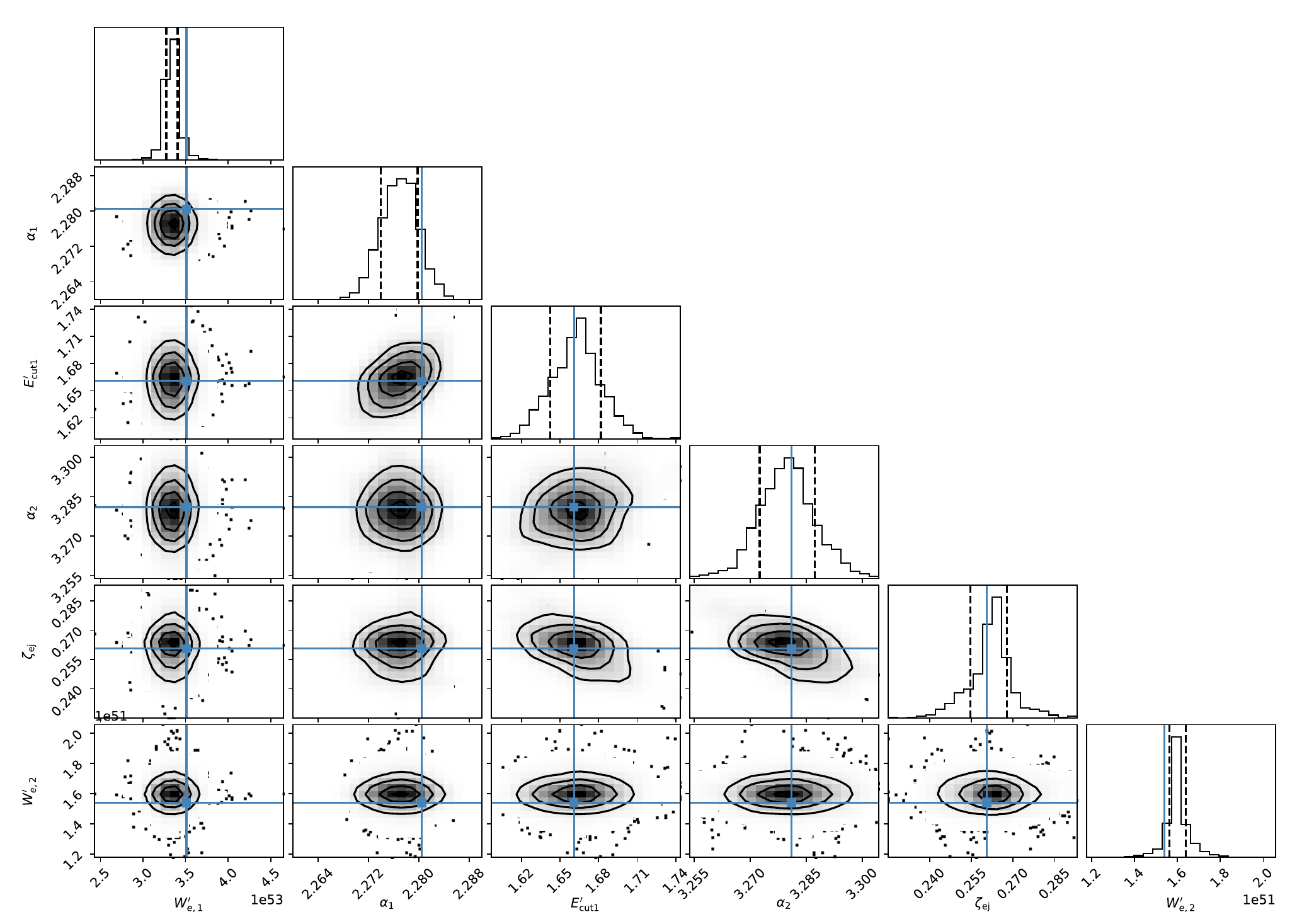} 
    \captionof{figure}{The corner image of the fitting parameters of knot A.} 
    \label{fig:corner}
\end{table*}



\renewcommand{\arraystretch}{1.1}
\begin{table*}
\caption{The \chandra\ Observations of M 87 with a frame time of 0.4 or 0.5.}
\centering
\begin{tabular}{cccc|cccc}
\hline\hline
ObsID & Frame Time & Exp Time & Start Date & ObsID & Frame Time & Exp Time & Start Date \\
& (s) & (ks) & (YYY-MM-DD) &&  (s) & (ks) & (YYY-MM-DD) \\ [0.1cm]
\hline
1808&0.4&12.85&2000-07-30&8578&0.4&4.71&2008-04-01 \\
3085&0.4&4.89&2002-01-16 &8579&0.4&4.71&2008-05-15 \\
3084&0.4&4.65&2002-02-12 &8580&0.4&4.7&2008-06-24\\
3086&0.4&4.62&2002-03-30 &8581&0.4&4.66&2008-08-07 \\
3087&0.4&4.97&2002-06-08 &10282&0.4&4.7&2008-11-17 \\
3088&0.4&4.71&2002-07-24 &10283&0.4&4.7&2009-01-07 \\
3975&0.4&5.29&2002-11-17&10284&0.4&4.7&2009-02-20\\
3976&0.4&4.79&2002-12-29 &10285&0.4&4.66&2009-04-01 \\
3977&0.4&5.28&2003-02-04 &10286&0.4&4.68&2009-05-13 \\
3978&0.4&4.85&2003-03-09 &10287&0.4&4.7&2009-06-22\\
3979&0.4&4.49&2003-04-14 &10288&0.4&4.68&2009-12-15 \\
3980&0.4&4.79&2003-05-18 &11512&0.4&4.7&2010-04-11 \\
3981&0.4&4.68&2003-07-03 &11513&0.4&4.7&2010-04-13 \\
3982&0.4&4.84&2003-08-08&11514&0.4&4.53&2010-04-15 \\
4917&0.4&5.03&2003-11-11 &11515&0.4&4.7&2010-04-17 \\
4918&0.4&4.68&2003-12-29 &11516&0.4&4.71&2010-04-20 \\
4919&0.4&4.7&2004-02-12 &11517&0.4&4.7&2010-05-05 \\
4921&0.4&5.25&2004-05-13 &11518&0.4&4.4&2010-05-09 \\
4922&0.4&4.54&2004-06-23 &11519&0.4&4.71&2010-05-11 \\
4923&0.4&4.63&2004-08-05 &11520&0.4&4.6&2010-05-14\\
5737&0.4&4.21&2004-11-26 &13964&0.4&4.54&2011-12-04 \\
5738&0.4&4.67&2005-01-24 &13965&0.4&4.6&2012-02-25\\
5739&0.4&5.15&2005-02-14 &14974&0.4&4.6&2012-12-12\\
5740&0.4&4.7&2005-04-22 &14973&0.4&4.4&2013-03-12\\
5744&0.4&4.7&2005-04-28 &16042&0.4&4.62&2013-12-26 \\
5745&0.4&4.7&2005-05-04 &16043&0.4&4.6&2014-04-02\\
5746&0.4&5.14&2005-05-13&17056&0.4&4.6&2014-12-17 \\
5747&0.4&4.7&2005-05-22 &17057&0.4&4.6&2015-03-19 \\
5748&0.4&4.7&2005-05-30 &18233&0.4&37.25&2016-02-23\\
5741&0.4&4.7&2005-06-03 &18781&0.4&39.51&2016-02-24\\
5742&0.4&4.7&2005-06-21 &18782&0.4&34.07&2016-02-26\\
5743&0.4&4.67&2005-08-06 &18809&0.4&4.52&2016-03-12\\
6299&0.4&4.65&2005-11-29 &18810&0.4&4.6&2016-03-13 \\
6300&0.4&4.66&2006-01-05 &18811&0.4&4.6&2016-03-14 \\
6301&0.4&4.34&2006-02-19 &18812&0.4&4.4&2016-03-16 \\
6302&0.4&4.7&2006-03-30 &18813&0.4&4.6&2016-03-17 \\
6303&0.4&4.7&2006-05-21 &18783&0.4&36.11&2016-04-20 \\
6304&0.4&4.68&2006-06-28 &18232&0.4&18.2&2016-04-27 \\
6305&0.4&4.65&2006-08-02 &18836&0.4&38.91&2016-04-28 \\
7348&0.4&4.54&2006-11-13 &18837&0.4&13.67&2016-04-30 \\
7349&0.4&4.68&2007-01-04&18838&0.4&56.29&2016-05-28 \\
7350&0.4&4.66&2007-02-13&18856&0.4&25.46&2016-06-12 \\ 
8510&0.4&4.7&2007-02-15&19457&0.4&4.60&2017-02-15 \\
8511&0.4&4.6&2007-02-18 &19458&0.4&4.58&2017-02-16 \\
8512&0.4&4.7&2007-02-21 &20034&0.4&13.12&2017-04-11 \\
8513&0.4&4.7&2007-02-24 &20035&0.4&13.12&2017-04-14 \\
8514&0.4&4.47&2007-03-12 &20488&0.4&4.6&2018-01-04 \\
8515&0.4&4.7&2007-03-14 &20489&0.4&4.6&2018-03-21 \\
8516&0.4&4.68&2007-03-19 &21075&0.4&9.13&2018-04-22 \\
8517&0.4&4.67&2007-03-22 &21076&0.4&9.04&2018-04-24 \\
7351&0.4&4.68&2007-03-24 &21457&0.4&14.12&2019-03-27 \\
7352&0.4&4.59&2007-05-15 &21458&0.4&12.76&2019-03-28 \\
7353&0.4&4.54&2007-06-25 &23669&0.4&13.67&2021-04-15\\
7354&0.4&4.71&2007-07-31 &23670&0.4&13.58&2021-04-16 \\
8575&0.4&4.68&2007-11-25&25302&0.5&4.71&2022-03-26\\ 
8576&0.4&4.69&2008-01-04 &25369&0.5&34.28&2022-05-06\\ 
8577&0.4&4.66&2008-02-16 &&&&\\
\hline
\end{tabular}
\label{table:chandra_obs}
\end{table*}

\bsp

\label{lastpage}
\end{document}